\documentclass[twocolumn,eqsecnum,aps,pra,superscriptaddress,10pt,showpacs]{revtex4-2}
\usepackage[usenames,dvipsnames]{color}
\usepackage[english]{babel}
\usepackage{epsfig}
\usepackage{graphicx,subfigure}
\usepackage{float}
\usepackage{tikz}
\usepackage{pgfplots}
\pgfplotsset{compat=1.12}
\usetikzlibrary{calc}
\usetikzlibrary{decorations.markings}
\usetikzlibrary{decorations.pathmorphing}
\usetikzlibrary{positioning,shapes,arrows,decorations.markings,calc}
\usepackage{longtable}
\usepackage{multirow}
\usepackage{arydshln}
\usepackage{hyperref}
\usepackage{amsmath,bbm,bm,amssymb}
\usepackage[mathcal,mathscr]{eucal}
\DeclareFontFamily{OT1}{pzc}{}
\DeclareFontShape{OT1}{pzc}{m}{it}{<-> s * [1.10] pzcmi7t}{}
\DeclareMathAlphabet{\mathpzc}{OT1}{pzc}{m}{it}
\usepackage[latin1]{inputenc}
\usepackage[T1]{fontenc}
\usepackage{times}
\usepackage{lmodern}
\usepackage[Euler]{upgreek}
\usepackage{enumitem}

\usepackage{bookmark}

\usepackage{textcomp}

\hypersetup{
  colorlinks=false,
  pdfborder={0 0 0},
}

\tikzset{midarrow/.style={decoration={
  markings,
  mark=at position #1 with {\arrow{angle 45}}},postaction={decorate}}
}
\tikzset{waved/.style={decorate,decoration=snake}}

\usepackage{soul}

\begin{document}
\setlength{\abovedisplayskip}{3pt}
\setlength{\belowdisplayskip}{3pt}
\setlength{\abovedisplayshortskip}{3pt}
\setlength{\belowdisplayshortskip}{3pt}

\renewcommand{\figurename}{FIG}
\renewcommand{\tablename}{TABLE}

\title{Structural disorder by octahedral tilting in inorganic halide perovskites: New insight with Bayesian optimization}

\author{Jingrui~Li*}
\email{jingrui.li@xjtu.edu.cn}
\affiliation{Electronic Materials Research Laboratory, Key Laboratory of the Ministry of Education and International Center for Dielectric Research, School of Electronic Science and Engineering \& International Joint Laboratory for Micro/Nano Manufacturing and Measurement Technology, Xi'an Jiaotong University, Xi'an 710049, China \looseness=-2}
\author{Fang~Pan}
\affiliation{Electronic Materials Research Laboratory, Key Laboratory of the Ministry of Education and International Center for Dielectric Research, School of Electronic Science and Engineering \& International Joint Laboratory for Micro/Nano Manufacturing and Measurement Technology, Xi'an Jiaotong University, Xi'an 710049, China \looseness=-2}
\author{Guo-Xu~Zhang}
\affiliation{MIIT Key Laboratory of Critical Materials Technology for New Energy Conversion and Storage, School of Chemistry and Chemical Engineering, Harbin Institute of Technology, Harbin 150001, China \looseness=-2}
\author{Zenghui~Liu}
\affiliation{Electronic Materials Research Laboratory, Key Laboratory of the Ministry of Education and International Center for Dielectric Research, School of Electronic Science and Engineering \& International Joint Laboratory for Micro/Nano Manufacturing and Measurement Technology, Xi'an Jiaotong University, Xi'an 710049, China \looseness=-2}
\author{Hua~Dong}
\affiliation{Key Laboratory for Physical Electronics and Devices of the Ministry of Education and Shaanxi Key Lab of Information Photonic Technique, School of Electronic Science and Engineering, Xi'an Jiaotong University, Xi'an 710049, China \looseness=-2}
\author{Dawei~Wang}
\affiliation{School of Microelectronics and Key Lab of Micro-Nano Electronics and System Integration of Xi'an City, Xi'an Jiaotong University, Xi'an 710049, China \looseness=-2}
\author{Zhuangde Jiang}
\affiliation{State Key Laboratory for Manufacturing Systems Engineering \& International Joint Laboratory for Micro/Nano Manufacturing and Measurement Technology, Xi'an Jiaotong University, Xi'an 710049, China \looseness=-2}
\author{Wei~Ren}
\affiliation{Electronic Materials Research Laboratory, Key Laboratory of the Ministry of Education and International Center for Dielectric Research, School of Electronic Science and Engineering \& International Joint Laboratory for Micro/Nano Manufacturing and Measurement Technology, Xi'an Jiaotong University, Xi'an 710049, China \looseness=-2}
\author{Zuo-Guang~Ye}
\affiliation{Department of Chemistry and 4D LABS, Simon Fraser University, Burnaby, British Columbia V5A 1S6, Canada \looseness=-2}
\author{Milica~Todorovi\'{c}*}
\email{milica.todorovic@utu.fi}
\affiliation{Department of Mechanical and Materials Engineering, University of Turku, FI-20014 Turku, Finland \looseness=-2}
\author{Patrick~Rinke}
\affiliation{Department of Applied Physics, Aalto University, P.O.Box 11100, FI-00076 AALTO, Finland \looseness=-2}

\begin{abstract}

Structural disorder is common in metal-halide perovskites and important for understanding the functional properties of these materials. First-principles methods can address structure variation on the atomistic scale, but they are often limited by the lack of structure-sampling schemes required to characterize the disorder. In this work, structural disorder in the benchmark inorganic halide perovskites $\text{CsPbI}_3^{}$ and $\text{CsPbBr}_3^{}$ is computationally studied in terms of the three octahedral-tilting angles. The consequent variation in energetics and properties are described by three-dimensional potential-energy surfaces (PESs) and property landscapes, delivered by Bayesian Optimization Structure Search method with integrated density-functional-theory (DFT) calculations. The rapid convergence of the PES with about 200 DFT data points in three-dimensional searches demonstrates the power of active learning and strategic sampling with Bayesian optimization. Further analysis indicates that disorder grows with increasing temperature, and reveals that the materials band gap at finite temperatures is a statistical mean over disordered structures.

\end{abstract}  

\maketitle

\section{Introduction}

Structural disorder in materials has become an important topic in both experimental and computational materials science \cite{Nayak2012,Cairns2013,Keen2015,Rhodes2019,Dragoe2019,Deringer2020,Simonov2020,Deringer2021}.  Disorder phenomena are very common in emergent functional materials, as ion-mixing or doping strategies are widely applied to obtain high performance and stability. Notable examples include ferroelectric oxide perovskite solid solutions \cite{YeZG2009,ZhangS2015,SunE2014}, multiple-cation metal halide perovskites for solar cells and light-emitting diodes \cite{AbdiJalebi2018a,CorreaBaena2019}, and kesterite photovoltaic materials \cite{ShinD2017,Giraldo2019}. Structural disorder can also occur in pure materials with perfect stoichiometry, if the material exhibits a series of stable structures with similar thermodynamic free energies, and the energy barriers separating free-energy minima are easy to overcome. Consequently, properties of complex functional materials often arise from the thermal population of a number of low-energy structures.

Cesium lead iodide ($\text{CsPbI}_3^{}$) is an example of the latter category that has received increasing attention in recent years. It is a promising photovoltaic material as $\text{CsPbI}_3^{}\,$-based perovskite solar cells have a reported power conversion efficiency above $21\%$ and good stability \cite{TanS2022}. It is generally believed that $\text{CsPbI}_3^{}$ adopts the cubic $\upalpha$ phase (space group $Pm\bar{3}m$) at high temperatures, which successively converts into the tetragonal $\upbeta$ phase ($P4/mbm$) at $533~\text{K}$ then the orthorhombic $\upgamma$ phase ($Pnma$) at $448~\text{K}$ upon cooling \cite{Stoumpos2015}. The structural details of the $\upbeta$ phase are not well established. In a joint experimental-theoretical study, Marronnier et~al. analyzed the phonon instabilities of this phase \cite{Marronnier2018}. Jinnouchi et~al. related the phase transition to the change in effective radius of $\text{Cs}^+$ by thermal fluctuations \cite{Jinnouchi2019}. Klarbring suggested that the macroscopic tetragonal phase consists of dynamically fluctuating orthorhombic structures \cite{Klarbring2019}. Yang et~al. further showed that some other low-symmetry structures are also involved \cite{ChenL2020}. Dynamical disorder has previously been found in other isostructural inorganic halide perovskites \cite{Patrick15,Klarbring2019,YangRX2020,ZhuX2022}. For example, a recent density-functional-theory (DFT) and molecular dynamics study indicated that the disorder in cubic $\text{CsPbBr}_3^{}$ is closely related to the octahedral-tilting dynamics \cite{ZhuX2022}. The nature of disorder in the high-temperature phases of inorganic halide perovskites remains unclear to-date, thus it is important to tackle this issue given the link between the atomic structure and functional properties of materials. Better knowledge about the involved structures, energetics, and mechanisms of structural fluctuations is crucial to understand the atomistic origin of functional properties and to design the next-generation high-performance perovskites.

To gain theoretical insight into structural disorder and its energetics, it is useful to study the system's multidimensional potential energy surface (PES) \cite{Kirkpatrick1983,Goedecker2005}, i.e., the total energy as a function of the relevant degrees of freedom (DOFs) for disorder. Energy differences governing structural disorder may be very small, so we must employ accurate DFT methods which have celebrated great successes in modeling molecules and condensed matter systems \cite{JonesRO2015}. DFT can supply one- and two-dimensional PESs with an equispaced-grid approach (such as in Ref.~\cite{Klarbring2019}). Yet PES computations quickly become intractable as the number of DOFs and thus the number of grid points grows. Attempts to construct accurate high-dimensional PESs based on first-principles calculations can be traced back several decades, when a series of interpolation schemes were developed to study polyatomic chemical reactions \cite{Bisseling1986,WuT2004}. However, it is not easy to directly apply this approach to complex-materials problems such as solid states and interfaces. Among the many atomic DOFs of a complex material, there are usually several ``principal'' DOFs that play a decisive role in structural energetics, while other DOFs are dependent or of secondary importance. As a result, such a PES could be much more complicated than that of a polyatomic molecular system. It might have multiple maxima and minima, making simple interpolation approaches difficult. 

In the past decade, machine learning (ML) has made an impact in the field of computational materials science \cite{Schleder2019b,Schmidt2019,Himanen2019,Deringer2019,Hart2021}. ML and DFT can be combined to approximate PESs \cite{Bartok2010,Behler2014} and to accelerate structure search with improved sampling schemes \cite{dAvezac2008,WangY2012,Bhattacharya2013,Yamashita2018,Joergensen2018} and accelerated force evaluation \cite{Bisbo2020,Mortensen2020,Kaappa2021,Wanzenboeck2022}. Force fields can of course provide the PESs in terms of all atomic DOFs yet typically lack the accuracy. Modern machine-learned force fields or interatomic potentials \cite{Rupp2012,Botu2017,vonLilienfeld2020,Sauceda2020,Vandermause2020,Timmermann2021,Unke2021,Westermayr2022} also provide DFT accuracy, but they take a long time to train in materials with many chemical species. We are here, however, interested in PESs that we can visualize, which is not possible for high-dimensional PESs of force fields.

Recently, some of us have developed the Bayesian Optimization Structure Search (BOSS) approach, an ML-based structure search scheme for accelerated and unbiased PES computation \cite{Todorovic2019}. BOSS couples state-of-the-art DFT calculations with the active learning Bayesian optimization (BO) technique. It employs Gaussian-process to fit a PES surrogate model to DFT data points, and refines this model until convergence is achieved by acquiring further data with a smart sampling strategy. In such a way, BOSS can construct a complete high-dimensional PES using a relatively modest number of energy data points. BOSS has already been applied to solve problems such as conformer search for organic molecules \cite{Todorovic2019,FangL2021} and adsorption of organic molecules at semiconductor surfaces \cite{Todorovic2019}, to resolve different organic adsorbate types and film growth at metallic surfaces \cite{EggerAT2020,Jaervi2020,Jaervi2021}, and to identify the interface geometries of inorganic or organic materials at perovskite surfaces \cite{Fangnon2022,ChenJ2022b,ZhangC2022}. The application of BO in materials science is not limited to structure search, e.g., it was used to train the force fields to study phase transition in hybrid perovskites \cite{Jinnouchi2019}, to select optimal ML hyperparameters \cite{Stuke2021,Laakso2022}, or to efficiently solicit experimental data \cite{SunS2021,Loefgren2022}.

In this work, we apply BOSS to study disorder in the benchmark systems $\text{CsPbI}_3^{}$ and the isostructural bromide perovskite ($\text{CsPbBr}_3^{}$). The latter is a standard green-emitting material for perovskite light-emitting diodes \cite{SimK2019}, another of today's key materials in perovskite optoelectronics. We select the three octahedral-tilting angles \cite{Glazer1972} as the principal DOFs. Previous experimental and theoretical studies \cite{ChungI2012,Stoumpos2013b,Patrick15,Klarbring2019,Gehrmann2019,ZhuX2022} have established that they are key to determining the materials structure. 
One of the main objectives of this work is 
to compute a realistic and accurate three-dimensional (3D) PES that best reflects the structures inside the perovskite materials. The smart data sampling by BOSS enables us to include all three octahedral-tilting angles as variables for the 3D PES. We thus go beyond the previously established two-dimensional (2D) view of Klarbring \cite{Klarbring2019}, which nevertheless is considered as reference for our study. In addition we use BOSS to generate a 3D profile of the important band gap property. This enables us to gain insight into the effects of structural disorder and temperature on functional properties. Important information about the structural disorder in both $\text{CsPbI}_3^{}$ and $\text{CsPbBr}_3^{}$ is obtained by gathering all these findings.

\section{Bayesian optimization structure search scheme}

\subsection{Octahedral tilting 
and the disorder search spaces}

\begin{figure*}[!ht]
\centering
\includegraphics[clip=true,trim=0.8in 5.6in 0.9in 0.7in,scale=0.96]{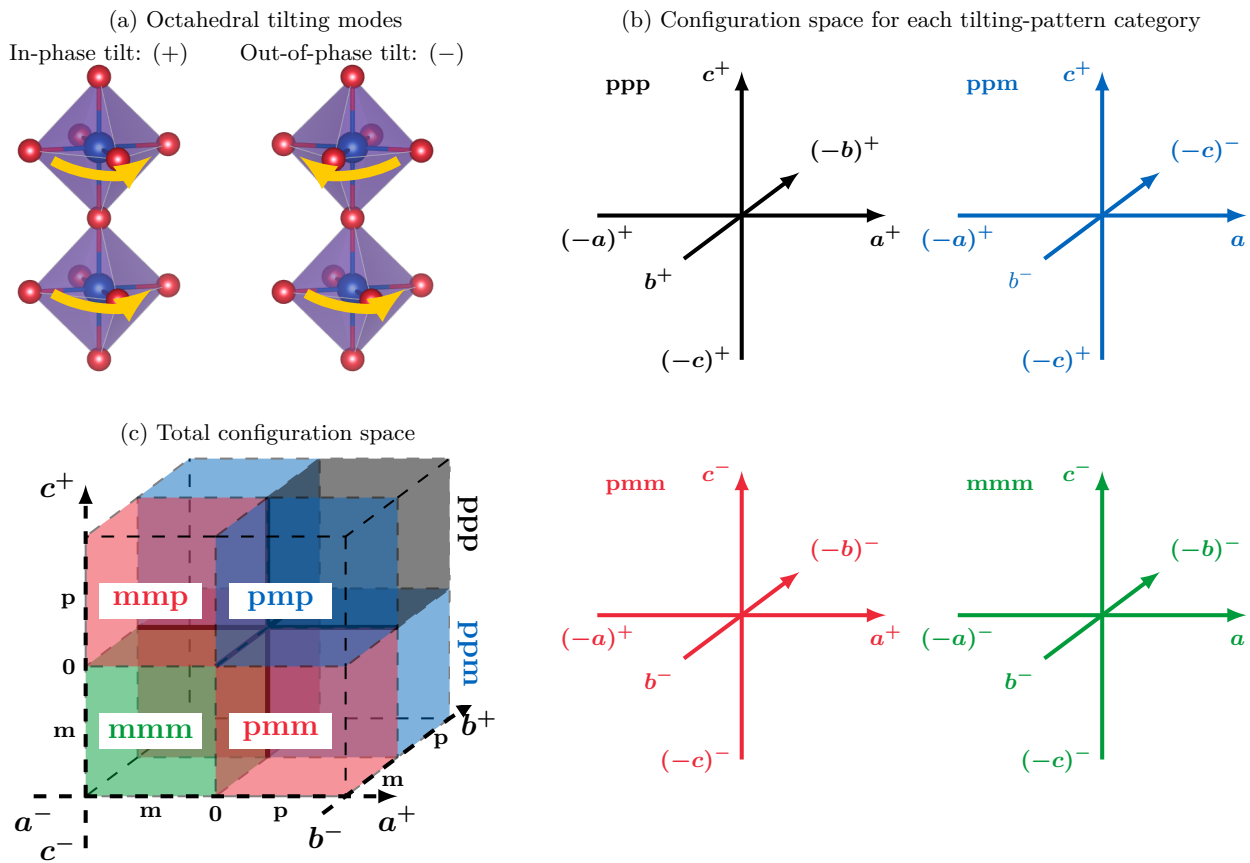}
\caption{Definition of octahedral-tilting configuration spaces for $\text{CsPbX}_3^{}$ ($\text{X}=\text{I},\text{Br}$) used in this work. (a) Tilting modes along a lattice vector: in-phase tilt (left, Glazer notation ``$+$'') and out-of-phase tilt (right, ``$-$''), with $\text{Pb}^{2+}$ cation colored in blue and $\text{X}^-$ anions in red, octahedral rotation directions indicated by yellow arrows; (b) the four configuration spaces representing different categories of octahedral-tilting patterns; and (c) the total configuration space. Spaces in (b) and octants  in (c) are colored to indicate the correspondence between them, e.g., the ``mmm'' space in (b) and the ``mmm'' octant in (c) are both colored in green.}
\label{fig:tilt}
\end{figure*}

The disorder search spaces for the benchmark systems $\text{CsPbI}_3^{}$ and $\text{CsPbBr}_3^{}$ are spanned by the three octahedral-tilting angles. We adopted the Glazer notation $a^{r_a^{}}b^{r_b^{}}c^{r_c^{}}$ of octahedral tilting in perovskites \cite{Glazer1972} throughout this manuscript. Here, $x^{r_x^{}}$ denotes the tilt around the lattice vector $x$ ($=a,b,c$) of the quasi-cubic cell, and $r_x^{}=+,-,0$ indicate the in-phase, out-of-phase, and zero tilting modes (sometimes shortened as ``tilts''), respectively (\textbf{Figure~\ref{fig:tilt}}a). The corresponding tilting angle is denoted by $\theta_x^{}$ hereafter. Different combinations of tilting modes and angles around the three lattice vectors result in different tilting patterns. The perovskite lattice accordingly exhibits different types of distortions. Originally, Glazer had derived 23 
tilting patterns \cite{Glazer1972,Glazer1975} and later Woodward revised this framework into 15 based on experimental results and space-group analysis \cite{Woodward1997a,Woodward1997b,Islam13}. In this work we considered all possible tilting patterns (listed in \textbf{Table~\ref{tab:sampling}}) that are compatible with both Glazer's and Woodward's conventions. A recent study of $\text{CsSnI}_3^{}\,$, which is isostructural and closely related to our systems, indicates that this is necessary \cite{XieN2020}. Results of DFT full relaxation for both systems with different tilting patterns are provided in Section~S1 in Supporting Information (SI).

\begin{table}[!ht]
\centering
\caption{Attribution of Glazer tilting patterns to different categories (disorder search spaces).}
\label{tab:sampling}
\vspace{-1em}
\begin{tabular}{lcccc}
& \multicolumn{1}{p{0.8cm}}{~} & \multicolumn{1}{p{0.8cm}}{~} & \multicolumn{1}{p{0.8cm}}{~} & \multicolumn{1}{p{0.8cm}}{~} \\ \hline
\multicolumn{1}{c}{Tilting} & ppp & ppm & pmm & mmm \\ \hline
$a^0a^0a^0$ & \checkmark & \checkmark & \checkmark & \checkmark \\
$a^+b^0b^0$ & \checkmark & \checkmark & \checkmark & \\
$a^-b^0b^0$ & & \checkmark & \checkmark & \checkmark \\
$a^+a^+c^0$, $a^+b^+c^0$ & \checkmark & \checkmark & & \\
$a^+b^-c^0$ & & \checkmark & \checkmark & \\
$a^-a^-c^0$, $a^-b^-c^0$ & & & \checkmark & \checkmark \\
$a^+a^+a^+$, $a^+a^+c^+$, $a^+b^+c^+$ & \checkmark & & & \\
$a^+a^+c^-$, $a^+b^+c^-$ & & \checkmark & & \\
$a^+b^-b^-$, $a^+b^-c^-$ & & & \checkmark & \\
$a^-a^-a^-$, $a^-a^-c^-$, $a^-b^-c^-$ & & & & \checkmark \\ \hline
\end{tabular}
\end{table}

The Glazer tilting patterns listed in Table~\ref{tab:sampling} can be divided into four categories: those with three in-phase tilts, with two in-phase and one out-of-phase tilts, with one in-phase and two out-of-phase tilts, and with three out-of-phase tilts. For each of them, the three octahedral-tilting angles as coordinates span a three-dimensional configuration space (Figure~\ref{fig:tilt}b). We denote them by $a^+b^+c^+$, $a^+b^+c^-$, $a^+b^-c^-$, and $a^-b^-c^-$, or simply by ppp, ppm, pmm, and mmm, respectively, with ``p'' standing for plus (``$+$'') and ``m'' for minus (``$-$'').
For each of them, the ``negative tilt'' $(-x)^{r_x^{}}$ describes the same tilt as the positive one, $(+x)^{r_x^{}}$ (i.e., $r_x^{}$ mode by $\theta_x^{}$ angle), but in the opposite direction \cite{Klarbring2019}. We should note two important features of these configuration spaces: (a) tilting patterns with zero tilt(s) belong to more than one spaces, e.g., $a^+b^+c^0$ belongs to both ppp and ppm, and (b) there are symmetry-equivalent points for each data (except $a^0a^0a^0$). As a result, any of the eight octants of a configuration space, e.g., octant~I with $\theta_a^{}\geqslant0,\theta_b^{}\geqslant0,\theta_c^{}\geqslant0$, contains all information for this category. A brief discussion about symmetry images of data entries that represent different tilting patterns are supplied in Section~S2 in SI. 

Figure~\ref{fig:tilt}c illustrates another way to construct the 3D disorder configuration space. All tilting patterns listed in Table~\ref{tab:sampling} are included in one space by assigning the positive and negative values of each tilting-angle coordinate to in-phase and out-of-phase tilts, respectively. We thus refer to it as the ``total configuration space'' hereafter. The color correspondence between Figure~\ref{tab:sampling}b and c indicates that each of the eight octants of this total space can represent a tilting-pattern category. For example, points in the mmm octant, i.e., $\theta_a^{}\leqslant0,\theta_b^{}\leqslant0,\theta_c^{}\leqslant0$, of this total space and points in the octant~I of the mmm space represent identical structures. In addition, either ppm or pmm has two symmetry-equivalent octants in the total space.

\subsection{BOSS protocol}

BOSS is an ML method that accelerates structure search via strategic sampling of the PES. In general, it can be used to rapidly explore $N$-dimensional domains and build surrogate models for any simulated properties. A more in-depth description of the BOSS scheme can be found in Refs.~\cite{Rasmussen2006,Gutman2016,Todorovic2019}. Here we only outline the search principle of BOSS with highlighting the aspects that are different to earlier applications.

First, the structures were sampled separately in the ppp, ppm, pmm, and mmm search spaces to ensure that there are enough data in all four tilting-pattern categories. The 3D surrogate models for visualization and analysis, in contrast, were constructed within the total space to avoid ambiguous data at the subspaces (planes and lines) shared by different tilting-pattern categories. 
Second, for each arrangement of tilting angles sampled, we performed a full structural relaxation of all other DOFs. We opted for this approach instead of the building-block models in previous BOSS applications to release any strains or structural artefacts introduced by the rearrangement of perovskite octohedra. 
Third, we guaranteed that the surrogate model was constructed and data acquisition was performed based on a set of DFT data points that reflect the structural symmetry of materials as alluded to. To this end, the sampling query additionally returned the data of all symmetry-equivalent points in the search space after each DFT calculation.

The BOSS workflow in this study is as follows. First, DFT calculations of tilted structures were separately sampled in each of the ppp, ppm, pmm, and mmm search spaces:
{\parindent 0em
\begin{enumerate}
[labelwidth=1.2em,leftmargin=1.2em,partopsep=0em,parsep=\parskip,itemsep=0em,topsep=0em]
  \item The database was initialized with three non-equivalent structures together with their symmetry images;
  \item A 3D surrogate model was fitted to all existing data using Gaussian process regression;
  \item Based on the surrogate model, an acquisition function was calculated and used to determine the next sample point in the configuration space;
  \item The database was updated with the newly sampled point and its symmetry images;
  \item Steps 2 \--- 4 were repeated until the parameters of the most-stable structure did not change within 10 consecutive iterations (variation of total energy within $0.5~\text{meV}$ per perovskite unit and all tilting angles within $0.5\text{\textdegree}$).
\end{enumerate}

Then the sampled data were merged into one data set according to the symmetry of the total configuration space. 3D surrogate models were fitted to both DFT-calculated total energy and band gap data to build the PES and band gap landscape, respectively. The surrogate model for the band gap was obtained in the same way from the same set of computations.
}

Once the PES was evaluated, we analyzed it to extract the locations of all local minima and their predicted energies. For all minima, we removed the fixed tilting angle restriction and optimized the structures fully to evaluate the quality of this small approximation. 
For any pair of minima, we utilized the surrogate model to compute the minimum energy path (MEP) between them and determine the energies and structures of transition states. 

For data sampling, we used the exploratory Lower Confidence Bound \cite{Brochu2010} acquisition function which balances exploitation (search for better points near the predicted global minimum) and exploration (acquiring data in the less visited regions of the search space). Such heuristic approaches are suitable for an unbiased exploration of the entire configuration space, not only limited to the vicinity of extrema.

For each sampled structure, 
we carried out DFT calculations for $\text{CsPbX}_3^{}$ using a $2\times2\times2$ supercell model that can host all Glazer tilting patterns. The $\text{Pb--X}$ framework of the system was initialized according to the desired tilting pattern. The $\text{Cs}$ locations and lattice parameters were relaxed during structural optimization, while the $\text{Pb}$ and $\text{X}$ fractional coordinates were frozen so that the tilting pattern was (at least approximately) maintained. We chose the Perdew-Burke-Ernzerhof exchange-correlation functional for solids (PBEsol) \cite{Perdew2008} and analytical stress tensor \cite{Knuth15} implemented in the all-electron numeric-atom-centered orbital code FHI-aims \cite{Blum09,HavuV09,RenX12,Levchenko15}, as PBEsol describes the lattice constants of halide perovskites well with moderate computational cost \cite{YangRX2017,Bokdam2017,Seidu2021a}. 
Single-point band gap calculations were performed for each optimized structure using a hybrid functional PBEsol0 that contains $25\%$ exact exchange. Scalar relativistic effects were included by means of the zero-order regular approximation \cite{vanLenthe93}, while spin-orbit coupling (SOC) was further included in PBEsol0 calculations. Standard FHI-aims tier-2 basis sets were used in combination with a $\Gamma$-centered $4\times4\times4$ $k$-point mesh.
The results of all relevant calculations of this work are available from the NOMAD (Novel Materials Discovery) repository \cite{NoMaD-BOSS_CPX}.

\section{Results and discussion}\label{results}

\subsection{Octahedral-tilting PESs}

\begin{figure*}[!ht]
\centering
\includegraphics[clip=true,trim=1.0in 4.89in 0.9in 0.7in,scale=0.96]{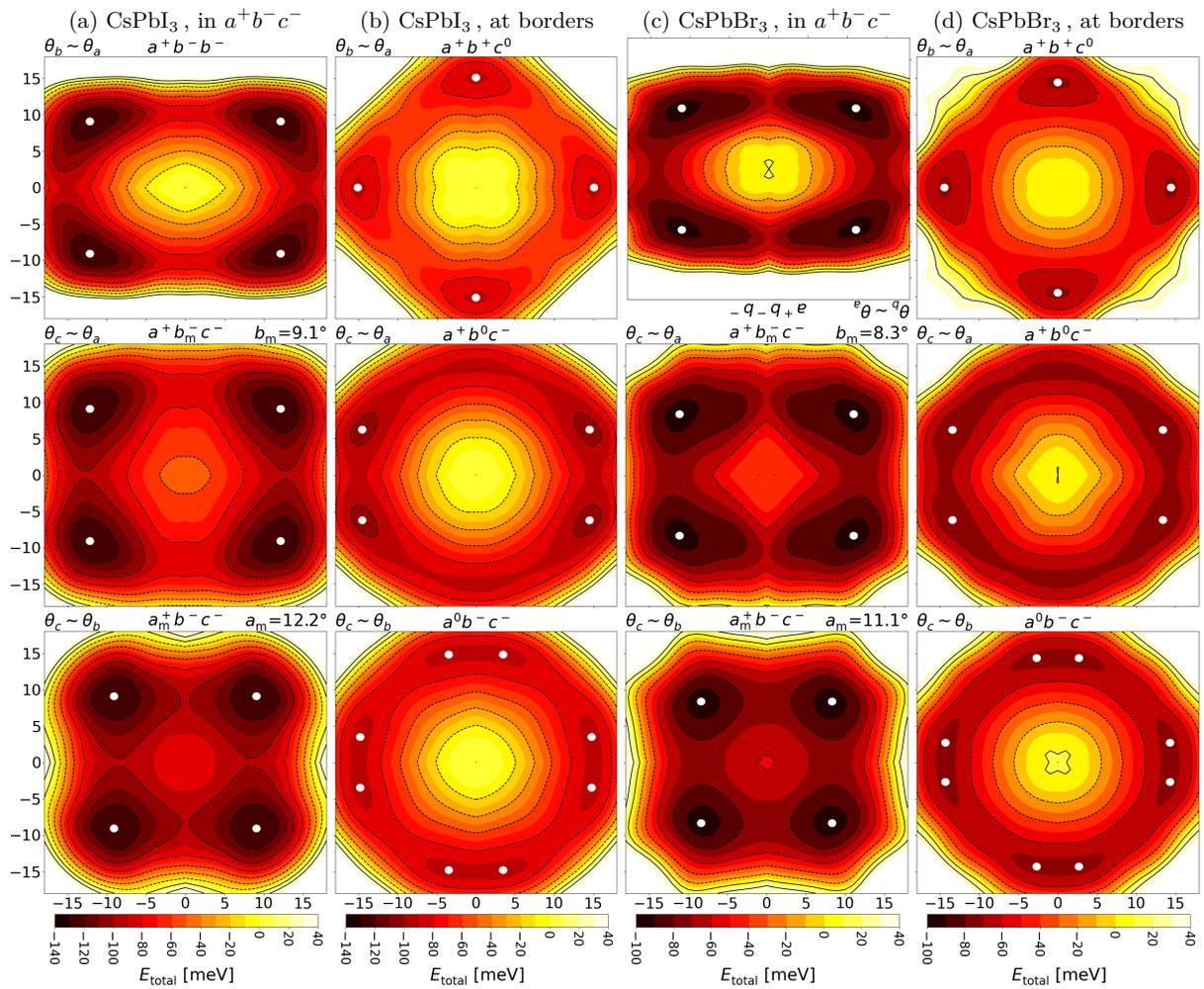}
\caption{2D cross-sections of 3D PESs of (a,b) $\text{CsPbI}_3^{}$ and (c,d) $\text{CsPbBr}_3^{}\,$. Total energies, $E_{\text{total}}^{}\,$, are given in their difference to the $a^0a^0a^0$ structure in $\text{meV}$ per $\text{CsPbX}_3^{}$ unit. For each material: (a,c) cross-sections $a^+b^-b^-$, $a^+b_{\text{m}}^-c^-$, and $a_{\text{m}}^+b^-c^-$ within pmm (subscript ``$\text{m}$'' in the Glazer tilting notation means that the corresponding variable is kept constant at its value at the PES minimum), with white dots indicating the global $E_{\text{total}}^{}$ minima in the total space; and (b,d) border planes $a^+b^+c^0$ (shared by ppp and ppm), $a^+b^0c^-$ (shared by ppm and pmm), and $a^0b^-c^-$ (shared by pmm and mmm), with white dots corresponding to the lowest-energies in ppp, ppm, and mmm, respectively. The axis names of each plot are given at the top-left corner, e.g., $\theta_b^{}\sim\theta_a^{}$ meaning $\theta_b^{}$ and $\theta_a^{}$ (both in $\text{\textdegree}$) for the vertical and horizontal axes, respectively. All figures have the same $x$ and $y$ scales. Note the different colormap scales for different materials.
}
\label{fig:CPX_PES}
\end{figure*}

The surrogate models for the energy landscapes produced by 3D octahedral tilting allow us insight into the energetics of disorder in $\text{CsPbI}_3^{}$ and $\text{CsPbBr}_3^{}\,$. 
We monitored the convergence of the models with structural sampling to ensure that the landscapes were well converged and reliable.  
\textbf{Figure~\ref{fig:CPX_PES}}a and b illustrate important 2D cross-sections of the 3D PES of $\text{CsPbI}_3^{}\,$, which was constructed by merging the first $23$, $75$, $76$, and $34$ DFT calculations ($809$, $795$, $797$, and $793$ symmetry-unfolded points) in the ppp, ppm, pmm, and mmm search spaces, respectively. The total-space database consists of $203$ non-equivalent DFT calculations which are symmetrized into $982$ data points. The 3D PES of $\text{CsPbBr}_3^{}$ (Figure~\ref{fig:CPX_PES}c and d) was fitted to $195$ DFT calculations ($937$ data points) collected from $38$, $65$, $70$, and $27$ DFT-calculation entries ($815$, $799$, $797$, and $793$ data points) in ppp, ppm, pmm, and mmm, respectively. Some further 2D cross-sections of the 3D PESs of both benchmark systems can be found in Section~S3 in SI.

At the outset, it is important to verify the quality of the PES surrogate models. 
The $a^+b^-b^-$ and $a^+b_{\text{m}}^-c^-$ cross-sections of $\text{CsPbI}_3^{}$ (Figure~\ref{fig:CPX_PES}a-top and middle, respectively) agree with the previous reported 2D PESs of the same material very well \cite{Klarbring2019}. We note that 
$76$ DFT calculations
sufficed to converge the 3D PES of pmm in our approach, while in Ref.~\cite{Klarbring2019}, $289$ DFT data points sampled over an equispaced grid in a quadrant of the 2D configuration space supplied a 2D PES. This comparison demonstrates the validity, accuracy, and the remarkable efficiency of our full-dimensional (3D) approach 
aided by Bayesian optimization.

The global minima (including symmetry-equivalent structures) in each search space are marked by white dots in Figure~\ref{fig:CPX_PES}. \textbf{Table~\ref{tab:CPX_min}} features the tilting angle locations and PES values of all minima, as well the results of subsequent full structure relaxations. 
The total energies derived from the surrogate model ($E_{\text{total}}^{}$) are only $<3.5~\text{meV}$ per perovskite unit higher than the DFT full-relaxation results ($E_{\text{f.r.}}^{}$), further confirming the accuracy of the 3D PES surrogate models. 
This difference is attributed to the constraint of the tilting patterns in DFT calculations. Since the difference is very small, we conclude that the three tilting angles are suitable principal DOFs. 

\begin{table}[!ht]
\centering
\caption{Structural data of global minima of all search spaces (the global minimum of the total space is marked by $^{\ast}$). Positive and negative tilting angles (in $\text{\textdegree}$) indicate in-phase and out-of-phase tilts, respectively. $E_{\text{total}}^{}$ is the total energy evaluated based on the 3D PES surrogate models. Also listed are the total energy of corresponding structures fully relaxed with DFT ($E_{\text{f.r.}}^{}$). Both energies are given in their difference to the $a^0a^0a^0$ structure in $\text{meV}$ per $\text{CsPbX}_3^{}$ unit.}
\label{tab:CPX_min} \vspace{-1em}
\begin{tabular}{crrrr}
& \multicolumn{1}{p{1.8cm}}{~} & \multicolumn{1}{p{1.8cm}}{~} & \multicolumn{1}{p{1.8cm}}{~} & \multicolumn{1}{p{1.8cm}}{~} \\ \hline
Search space &
\multicolumn{1}{c}{${\color{white}{^{\ast}}}$pmm$^{\ast}$} &
\multicolumn{1}{c}{ppp} & \multicolumn{1}{c}{ppm} & \multicolumn{1}{c}{mmm} \\ \hline
\multicolumn{5}{l}{$\text{CsPbI}_3^{}$} \\
$\theta_a^{}$
&   $12.2$~~~ &   $15.1$~~~ &   $14.5$~~~ &  $-14.8$~~~ \\
$\theta_b^{}$
&   $-9.1$~~~ & $0{\color{white}{.0}}$~~~ & $0{\color{white}{.0}}$~~~
                                          &   $-3.5$~~~ \\
$\theta_c^{}$
&   $-9.1$~~~ & $0{\color{white}{.0}}$~~~ &   $-6.2$~~~ & $0{\color{white}{.0}}$~~~ \\
$E_{\text{total}}^{}$
& $-129.2$~~~ &  $-91.3$~~~ & $-103.5$~~~ &  $-93.2$~~~ \\
Space group
& \multicolumn{1}{c}{$Pnma$}
& \multicolumn{1}{c}{$P4/mbm$}
& \multicolumn{1}{c}{$Cmcm$}
& \multicolumn{1}{c}{$C2/m$} \\
$E_{\text{f.r.}}^{}$
& $-132.6$~~~ &  $-91.5$~~~ & $-104.1$~~~ &  $-93.7$~~~ \\
In Figure~\ref{fig:CPX_PES}
& \multicolumn{1}{c}{(a)}
& \multicolumn{1}{c}{(b)-$a^+b^+c^0$}
& \multicolumn{1}{c}{(b)-$a^+b^0c^-$}
& \multicolumn{1}{c}{(b)-$a^0b^-c^-$} \vspace{0.5em} \\ \hline
\multicolumn{5}{l}{$\text{CsPbBr}_3^{}$} \\
$\theta_a^{}$
&   $11.1$~~~ &   $14.4$~~~ &   $13.4$~~~ &  $-14.3$~~~ \\
$\theta_b^{}$
&   $-8.3$~~~ & $0{\color{white}{.0}}$~~~ & $0{\color{white}{.0}}$~~~
                                          &   $-2.7$~~~ \\
$\theta_c^{}$
&   $-8.3$~~~ & $0{\color{white}{.0}}$~~~ &   $-6.2$~~~ & $0{\color{white}{.0}}$~~~ \\
$E_{\text{total}}^{}$
&  $-93.9$~~~ &  $-71.5$~~~ &  $-77.5$~~~ &  $-72.5$~~~ \\
Space group
& \multicolumn{1}{c}{$Pnma$}
& \multicolumn{1}{c}{$P4/mbm$}
& \multicolumn{1}{c}{$Cmcm$}
& \multicolumn{1}{c}{$C2/m$} \\
$E_{\text{f.r.}}^{}$
&  $-97.3$~~~ &  $-71.5$~~~ &  $-78.3$~~~ &  $-73.2$~~~ \\
In Figure~\ref{fig:CPX_PES}
& \multicolumn{1}{c}{(c)}
& \multicolumn{1}{c}{(d)-$a^+b^+c^0$}
& \multicolumn{1}{c}{(d)-$a^+b^0c^-$}
& \multicolumn{1}{c}{(d)-$a^0b^-c^-$} \\ \hline
\end{tabular}
\end{table}

The lowest-energy structures of both materials exhibit an $a^+b^-b^-$ tilting ($\upgamma$ phase, space group $Pnma$) which is found in pmm (Figure~\ref{fig:CPX_PES}a and c). The 2D cross-sections of the 3D PESs in this space are qualitatively similar for these two materials. The main difference is that the global minimum of $\text{CsPbI}_3^{}$ is lower than $\text{CsPbBr}_3^{}$ by $35.3~\text{meV}$ per unit. In addition, we can observe less steep low-energy basins with $\text{CsPbBr}_3^{}$ than  $\text{CsPbI}_3^{}\,$. 
This can be rationalized by structural consideration: the Goldschmidt tolerance factor of $\text{CsPbBr}_3^{}$ ($0.862$) is larger than that of $\text{CsPbI}_3^{}$ ($0.851$) due to the smaller radius of the $\text{Br}^-$ anion. $\text{CsPbI}_3^{}$ therefore has a stronger tendency of octahedral tilting to stabilize the structure, resulting in more energy gain compared to the untilted structure and a steeper PES in the vicinity of the minimum.

Based on the 3D PES surrogate models, we can use the nudged elastic band (NEB) method to easily evaluate the barrier of the MEPs that connect a global minimum and its symmetry image. Important results are presented in \textbf{Table~\ref{tab:CPX_barrier}} (see Section~S4 of SI for more detail). The smallest barriers are $25.7$ and $16.4~\text{meV}$ for $\text{CsPbI}_3^{}$ and $\text{CsPbBr}_3^{}\,$, respectively, associated with the $a^+b^-b^- \rightarrow a^+b^-c^0 \rightarrow a^+b^-(-b)^-$ transition path. The lower barrier of $\text{CsPbBr}_3^{}$ is a direct result of the shallower basin in the vicinity of the global minima. The MEPs linking $a^+b^-b^-$ and $(-a)^+b^-b^-$ have higher barriers, indicating that larger energy is required for reversing the direction of an in-phase tilt than for an out-of-phase tilt. 
The transition $a^+b^-b^- \rightarrow a^+(-b)^-(-b)^-$ follows two consecutive MEPs, each associated with the direction reversal of an out-of-phase tilt. A simultaneous direction reversal of the two out-of-phase tilts is not favored. Our findings for $\text{CsPbI}_3^{}$ that (a) the lowest barrier is associated with a transition-state structure with an $a^+b^-c^0$ tilting, and (b) its energy is lower than what is found in the 2D approach 
by a few $\text{meV}$ (see Section~S4 of SI), are in good agreement with Klarbring's results \cite{Klarbring2019}. The advantage of our BOSS approach is that we can directly evaluate the MEP within the full-dimensional (3D) PES, so that we neither need to calculate several 2D PESs nor will miss the path with the lowest barrier.

\begin{table*}[!ht]
\centering
\caption{MEPs connecting the global-minimum structure $a^+b^-b^-$ and its symmetry equivalent structures and the associated barriers.
} \label{tab:CPX_barrier} \vspace{-1em}
\begin{tabular}{l@{\hspace{2em}}cc}
\multicolumn{1}{p{2cm}}{~} & \multicolumn{1}{p{1.4cm}}{~} & \multicolumn{1}{p{1.4cm}}{~} \\ \hline
& \multicolumn{2}{c}{Barrier $[\text{meV}]$} \\
\multicolumn{1}{c}{\multirow{-2}{*}{Structure-variation MEP}} & $\text{CsPbI}_3^{}$ & $\text{CsPbBr}_3^{}$ \\ \hline
$a^+b^-b^- \rightarrow a^+b^-c^0 \rightarrow a^+b^-(-b)^-$ & $25.7$ & $16.4$ \\
$a^+b^-b^- \rightarrow a^0b^-c^- \rightarrow (-a)^+b^-b^-$ & $35.9$ & \\
$a^+b^-b^- \rightarrow a^0b^-b^- \rightarrow (-a)^+b^-b^-$ & & $23.1$ \\
$a^+b^-b^- \rightarrow a^+b^-c^0 \rightarrow a^+b^-(-b)^- \rightarrow a^+b^0(-c)^- \rightarrow a^+(-b)^-(-b)^-$ & $25.7$ & $16.4$ \\ \hline
\end{tabular}
\end{table*}

The other three search spaces are naturally of less interest than pmm as they do not contain the overall most stable structure. For an in-depth insight of the tilting modes, we nevertheless briefly discuss them based on Figure~\ref{fig:CPX_PES}b and d. Table~\ref{tab:CPX_min} already indicates that the white dots in the top, middle, and bottom of Figure~\ref{fig:CPX_PES}b or d are the global minima in the ppp, ppm, and mmm search spaces, respectively. The lowest-energy structure of ppp has an $a^+b^0b^0$ tilting (space group $P4/mbm$) which is generally considered as the structure of the recently reported tetragonal $\upbeta$ phase \cite{Stoumpos2015,Marronnier2018,WangY2019}. This structure is very localized in terms of the single in-phase tilting angle, which is 
obviously larger than the in-phase tilting angle of the overall most stable ($a^+b^-b^-$) structure (see Table~\ref{tab:CPX_min}). This structure also belongs to ppm but is not the most stable one within this search space. The global minimum of ppm is instead located at $a^+b^0c^-$ whose energy is lower than the $a^+b^0b^0$ structure by $12.2~\text{meV}$. It is the transition-state structure of the $a^+b^-b^- \rightarrow a^+(-b)^-b^-$ MEP as alluded to. The local minimum of mmm exhibits an $a^-b^-c^0$ tilting with one large ($>10\text{\textdegree}$) and one small ($<5\text{\textdegree}$) out-of-phase tilt. The barrier of the amplitude-exchange pathway, $a^0b^-c^- \rightarrow a^0c^-b^-$, is rather low (see Section~S4 of SI). Finally, none of the $a^-b^0b^0$ (space group $I4/mcm$) structures 
is a global minimum of any search space. This might explain why it has never been experimentally observed. 

Summarizing Figures~\ref{fig:CPX_PES}b and d provides insight of structural disorder in these two benchmark systems. The in-phase and out-of-phase tilts have different characters. An in-phase tilt around one lattice vector ``excludes'' in-phase tilts around any other lattice vectors; while an out-of-phase tilt can stabilize the structure with an in-phase tilt, and can co-exist and exchange energy with another out-of-phase tilt. We can thus conclude that the out-of-phase tilting mode is more active than the in-phase mode in $\text{CsPbX}_3^{}\,$. The motion of out-of-phase tilts transfers vibrational energy from one lattice vector to another, giving rise to noticeable dynamical structural disorder.

\subsection{Configuration space distribution of structures at finite temperatures}

\begin{figure}[!ht]
\centering
\includegraphics[scale=0.36]{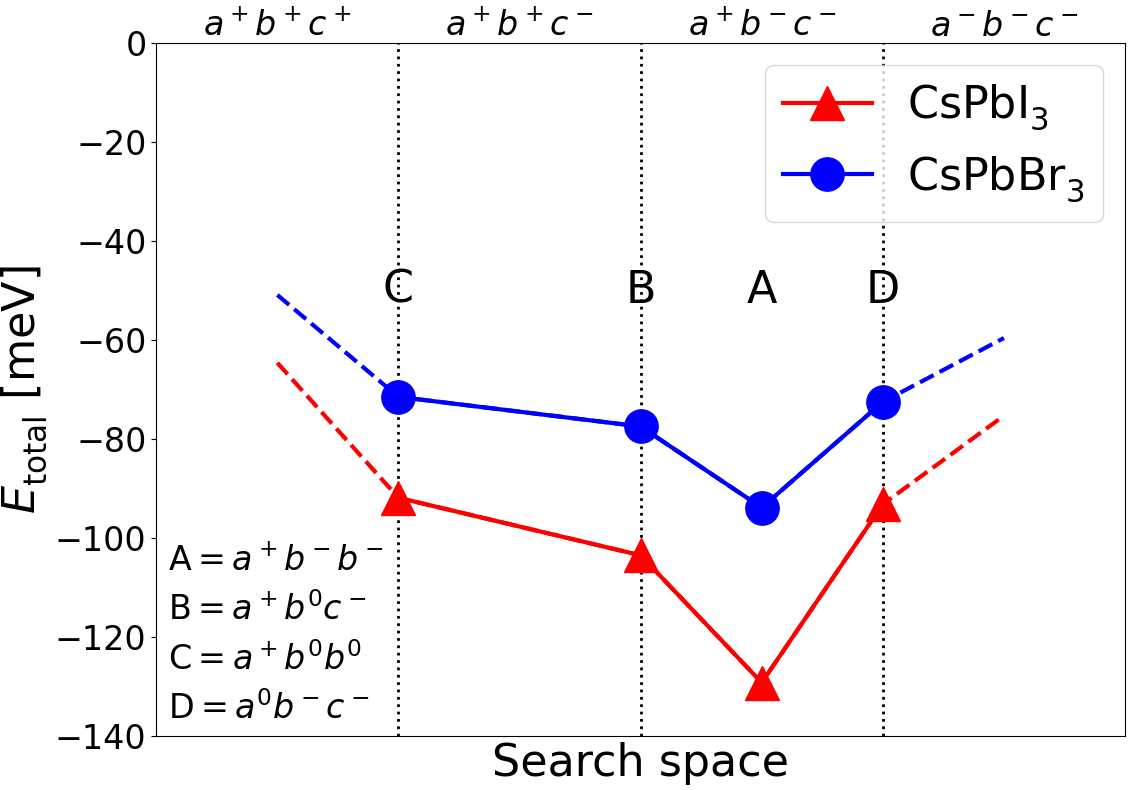}
\caption{Energies of the most stable structures (labeled by A\---D) in the four search spaces (red triangles and blue circles for $\text{CsPbI}_3^{}$ and $\text{CsPbBr}_3^{}\,$, respectively).}
\label{fig:energy}
\end{figure}

Based on the total-energy profiles in the full-dimensional (3D) configuration space, we can analyze the distribution of the structural disorder of $\text{CsPbI}_3^{}$ and $\text{CsPbBr}_3^{}$ at finite temperatures. \textbf{Figure~\ref{fig:energy}} sketches the total energy data listed in Table~\ref{tab:CPX_min} and facilitates the prediction of how structures evolve and disorder grows in both systems with increasing temperature.

\begin{figure*}[!ht]
\centering
\includegraphics[clip=true,trim=1.0in 4.9in 0.9in 0.7in,scale=0.96]{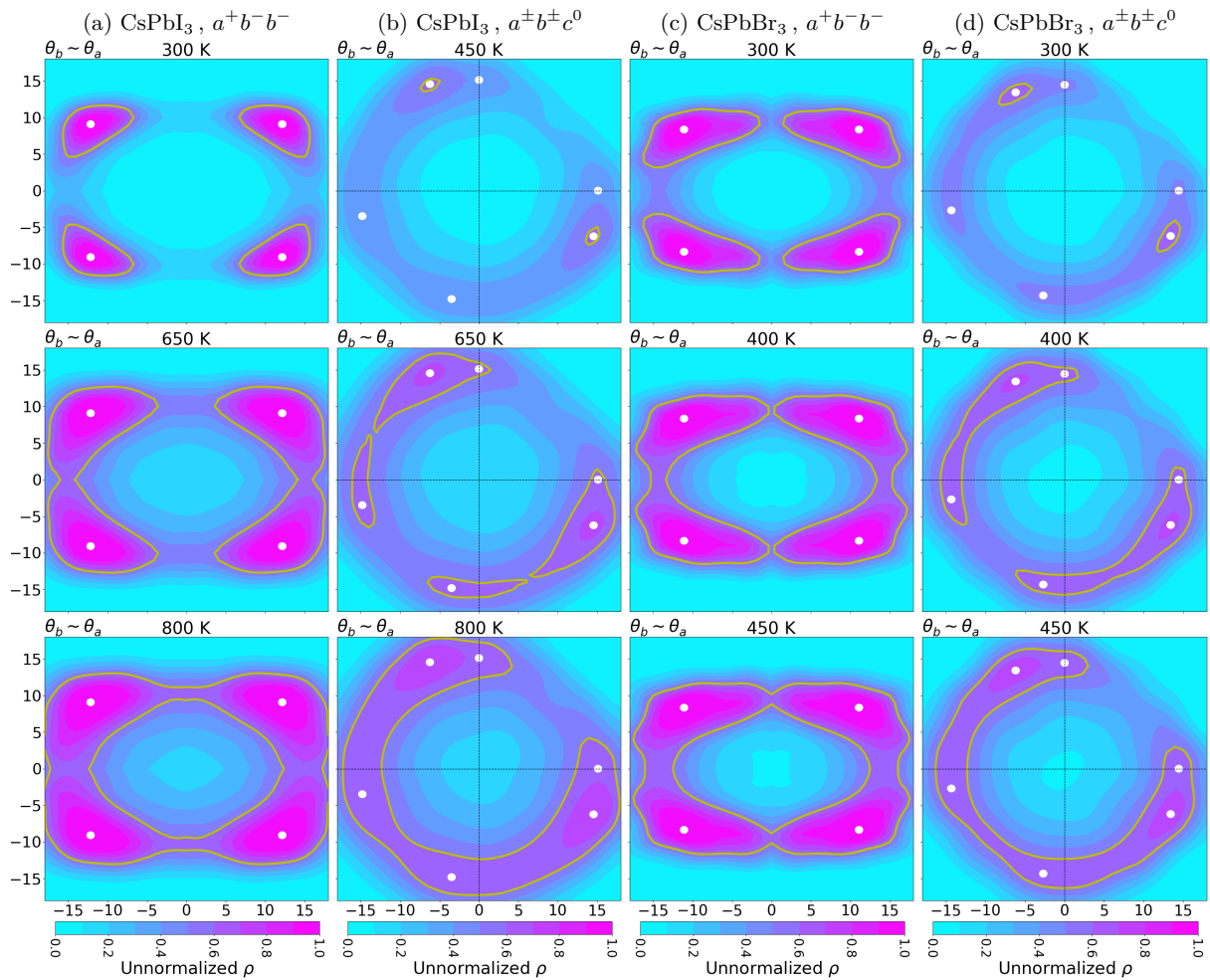}
\caption{2D cross-sections of unnormalized distribution probability density of (a,b) $\text{CsPbI}_3^{}$ and (c,d) $\text{CsPbBr}_3^{}$ in the search spaces of tilting angles at different temperatures. Shown for each material: (a,c) $a^+b^-b^-$ cross-section from pmm, and (b,c) $a^{\pm}b^{\pm}c^0$ cross-section from the total configuration space. The thick yellow lines in each plot mark a relative probability of $0.5\,$.}
\label{fig:thermal}
\end{figure*}

From the BOSS surrogate model of PES (Figure~\ref{fig:CPX_PES}), we calculated the probability density of finding $\text{CsPbI}_3^{}$ or $\text{CsPbBr}_3^{}$ in a structure that is labeled by point $\bm{\theta}$ (a shortened notation of tilting angles) in the total configuration space at temperature $T$ using the Boltzmann distribution
\begin{align*}
\rho(\bm{\theta},T) &\propto \text{e}^{-[E(\bm{\theta})-E_{\min}^{}]/k_{\text{B}}^{}T}\,.
\end{align*}
Here $E(\bm{\theta})$ and $E_{\min}^{}$ denote the energies of structure $\bm{\theta}$ and the overall most stable structure, respectively. $k_{\text{B}}$ is the Boltzmann constant. In this work, we have not normalized the total probability but rather kept the maximum probability density always at $1$. We can thus set a criterion of $\rho$ ($0.5$ in this work) to help investigate the distribution of thermal population of disordered structures. We are especially interested in two features: (a) the region within which $\rho>0.5$ (the larger the region, the more extensive the disorder), and (b) whether the $\rho>0.5$ distribution emerges at some particular regions in the configuration space (meaning noticeable population at the corresponding tilting patterns). Representative results are illustrated in \textbf{Figure~\ref{fig:thermal}} (see also Section~S5 of SI).

Figure~\ref{fig:thermal}a shows that $\text{CsPbI}_3^{}$ is quite strongly bound in its most stable $a^+b^-b^-$ ($Pnma$, $\upgamma$ phase) structure at $300~\text{K}$. As temperature increases, the $\rho>0.5$ regions grow gradually and reach the $a^+b^-c^0$ plane first at $450~\text{K}$ (Figure~\ref{fig:thermal}b). This corresponds to the A $\rightarrow$ B transition in Figure~\ref{fig:energy}, which requires an energy of $25.7~\text{meV}$. At this point, the structure has not yet gained enough energy to further travel across a series of $a^+b^+c^-$ structures (B $\rightarrow$ C in Figure~\ref{fig:energy}), but will rather be ``reflected'' by this $a^+b^-c^0$ border back to pmm. It can either go back to the initial $a^+b^-b^-$ structure, or fall into another equivalent $a^+b^-(-b)^-$ minimum with the direction of one out-of-phase tilt reversed around the lattice vector. Our observation of the $a^+b^-b^-\rightarrow a^+b^-c^0\rightarrow a^+b^-(-b)^-$ mechanism is in good agreement with the previous study of the structural disorder for the observed $\upbeta$-$\text{CsPbI}_3^{}$ by Klarbring \cite{Klarbring2019}.

As temperature further increases beyond $450~\text{K}$, noticeable population of $\text{CsPbI}_3^{}$ structure at $a^+b^0b^0$ (Figure~\ref{fig:thermal}a) and $a^-b^-c^0$ (Figure~\ref{fig:thermal}b) is observed at $650~\text{K}$. The former corresponds to ``C'' in Figure~\ref{fig:energy}, with only one single in-phase tilt while both out-of-phase tilts are largely suppressed. The latter corresponds to ``D'' in Figure~\ref{fig:energy}. The structure can easily fluctuate between in-phase and out-of-phase tilts  at this temperature, too. The direction reversal of the in-phase tilt, $a^+\rightarrow(-a)^+$, becomes highly probable at $800~\text{K}$. The motion of the less active in-phase tilt gives rise to an extremely high level of structural disorder.

$\text{CsPbBr}_3^{}$ exhibits generally similar features (Figure~\ref{fig:thermal}c and d), yet noticeable differences can be observed. Because of the shallower PES basins, the probability distribution of $\text{CsPbBr}_3^{}$ at $300~\text{K}$ is much broader than $\text{CsPbI}_3^{}\,$. First, $\rho>0.5$ can already be observed at $a^+b^-c^0$ at room temperature due to a much lower activation energy ($16.4~\text{meV}$). At $400~\text{K}$,  the $a^+b^0b^0$ ($P4/mbm$) structure is populated. At this temperature, we also observe $a^+b^0\leftrightarrow a^0b^-$ tilting-mode transitions. Finally, at $450~\text{K}$ the $a^+\leftrightarrow(-a)^+$ motion sets in. All of these findings indicate a much more disordered structure in $\text{CsPbBr}_3^{}$ than its iodide analog at the same temperature.

\textbf{Table~\ref{tab:CPX_thermal}} summarizes the important phenomena derived from Figure~\ref{fig:thermal} and Figure~S2. Both benchmark systems exhibit a three-stage mechanism of structure variation as temperature increases. Starting from the most stable $a^+b^-b^-$ structure, we can first observe the motion of one out-of-phase tilting mode, $a^+b^-b^- \leftrightarrow a^+b^-c^0$. This process is closely related to the $\upgamma \rightarrow \upbeta$ phase transition. In our evaluation, the temperature of this stage is very close to the phase-transition temperature for $\text{CsPbI}_3^{}$ ($450$ vs. $448~\text{K}$ \cite{Stoumpos2015,Marronnier2018}) but clearly lower than that for $\text{CsPbBr}_3^{}$ ($300$ vs. $361~\text{K}$ \cite{Hirotsu1974,Stoumpos2013b}). Further increase of temperature activates the motion of another out-of-phase tilt and the fluctuation between the out-of-phase and in-phase tilts. Finally, direction reversal of any mode and switch between any two modes become prevalent, giving rise to the highly disordered, effectively cubic structure with pronounced lattice vibration. Our evaluation of this stage's temperature is a bit higher than the $\upbeta \rightarrow \upalpha$ phase-transition temperature for $\text{CsPbBr}_3^{}$ ($450$ vs. $403~\text{K}$ \cite{Hirotsu1974,Stoumpos2013b}) and much higher than that for $\text{CsPbI}_3^{}$ ($800$ vs. $533~\text{K}$ \cite{Stoumpos2015,Marronnier2018}). The focus of this work is to understand the nature of structural disorder of these emergent and important optoelectronic materials at the atomic scale. The PES database created in this paper has laid a solid groundwork for the direct simulation of the $\upgamma\rightarrow\upbeta\rightarrow\upalpha$ phase transition, which is important but beyond the scope of this work.

\begin{table}[!ht]
\centering
\caption{Important tilting patterns: the (approximate) temperature at which they become noticeably populated, and the tilting-mode transition to which they closely correspond.}
\label{tab:CPX_thermal} \vspace{-1em}
\begin{tabular}{clcc}
& & \multicolumn{1}{p{1.4cm}}{~} & \multicolumn{1}{p{1.4cm}}{~} \\ \hline
& \multicolumn{1}{c}{Relevant transition} & \multicolumn{2}{c}{$T$ of emergence $[\text{K}]$} \\
\multirow{-2}{*}{Tilting pattern} & \multicolumn{1}{c}{of tilting modes} & $\text{CsPbI}_3^{}$ & $\text{CsPbBr}_3^{}$ \\ \hline
$a^+b^-c^0$ & $a^+b^-b^-\rightarrow a^+b^-(-b)^-$ & $450$ &   $<300$ \\
$a^+b^0b^0$ & $\,\,a^+b^-c^0\rightarrow a^+(-b)^-c^0$ & $650$ & $~~~400$ \\
$a^-b^0b^0$ & $\,\,\,a^+b^0b^0\rightarrow a^0b^-a^0$  & $650$ & $~~~400$ \\
$a^-a^-c^0$ & $\,\,\,a^-b^0b^0\rightarrow a^0b^-a^0$  & $800$ & $~~~450$ \\ \hline
\end{tabular}
\end{table}

As a concluding note, we emphasize that the zero-tilting structure $a^0a^0a^0$, which is generally regarded as the $\upalpha$ (cubic) phase, is very rarely populated even at $>800~\text{K}$. This is simply because it is energetically much less favorable than the most stable $a^+b^-b^-$ structure. Our results indicate that the cubic phase must be understood as a dynamical mixture of low-symmetry structures, a result that was found by a recent DFT and molecular dynamics study for $\text{CsPbBr}_3^{}$ \cite{ZhuX2022}. In this phase, plenty of domains or nano-regions of low-symmetry structures (such as $a^+b^-b^-$) occur with equal probability; the octahedra of each nano-region are dynamically fluctuating to be adaptive to their neighbors, thus giving rise to an effectively cubic average structure.

\subsection{Band gap distributed over disordered structures}

With the sampled DFT calculations we are able to fit surrogate models for materials properties of interest other than PESs. The band gap is one of the most important electronic-structure properties for optoelectronic materials. Here we focus on how band gaps of the benchmark systems vary with octahedral-tilting angles.

\begin{figure*}[!ht]
\centering
\includegraphics[clip=true,trim=0.7in 5.7in 0.7in 0.7in,scale=0.925]{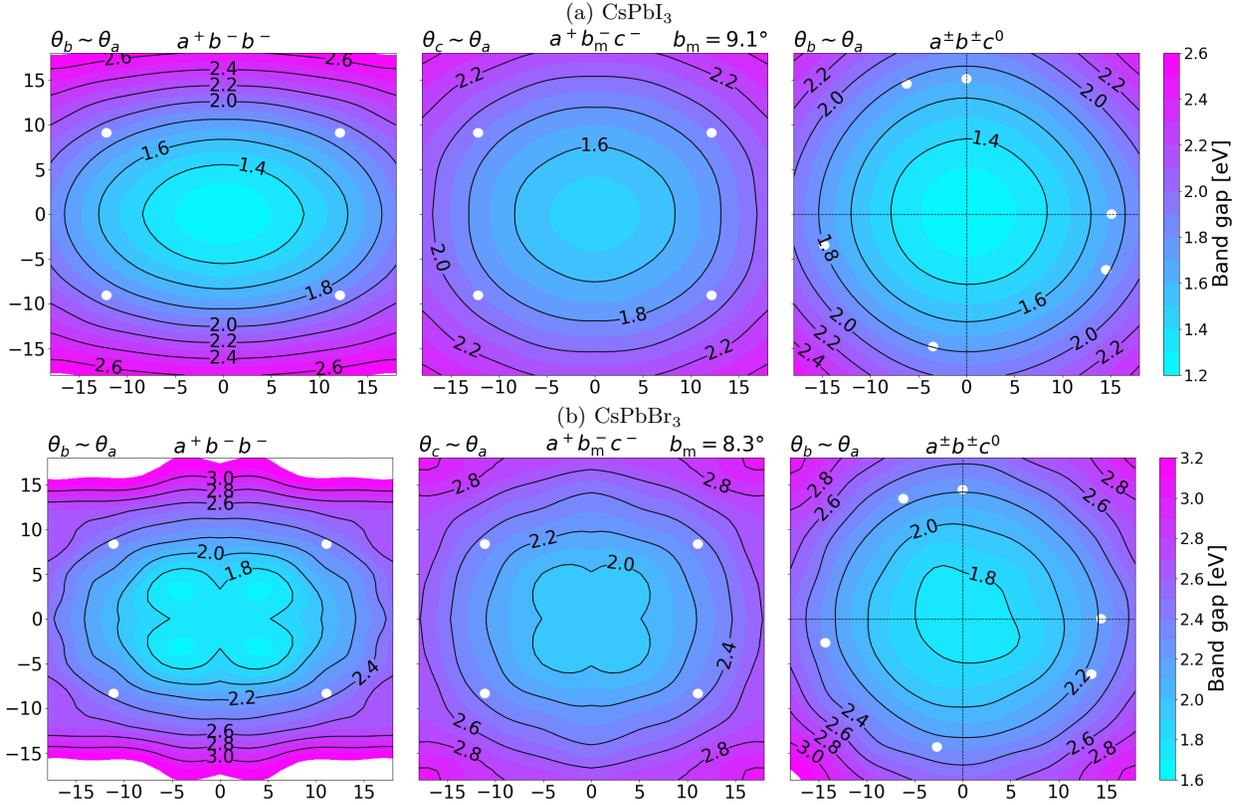}
\caption{Band gaps of (a) $\text{CsPbI}_3^{}$ and (b) $\text{CsPbBr}_3^{}$ in the search space of tilting angles (shown are 2D cross-sections $a^+b^-b^-$, $a^+b_{\text{m}}^-c^-$, and $a^{\pm}b^{\pm}c^0$ of the 3D surrogate models).
}
\label{fig:bands}
\end{figure*}

\textbf{Figure~\ref{fig:bands}} provides representative 2D cross-sections of 3D surrogate models that are fitted to the band gap data calculated with DFT (PBEsol0+SOC). For $\text{CsPbI}_3^{}\,$, we observe that the band gap increases monotonically with tilting angle. While for $\text{CsPbBr}_3^{}$ the overall trend is the same, small deviations are observed when all three tilting angles are small.

It it noteworthy that the cubic $a^0a^0a^0$ structure has an obviously lower band gap than all minima structures in different search spaces (white dots in Figure~\ref{fig:bands}) for each benchmark system. This is due to the maximal overlap of atomic orbitals as a result of the zero tilt, which broadens the valence and conduction bands thus reducing the gap. As already alluded to, such a cubic structure is not seen in experiments, because it is too high in energy. The experimental band gaps of $\text{CsPbI}_3^{}$ and $\text{CsPbBr}_3^{}$ are $\sim1.7$ \cite{Eperon14,Eperon15,Stoumpos2015} and $\sim2.3~\text{eV}$ \cite{Stoumpos2013b,Mannino2020}, respectively. DFT band gaps of all minima structures fall into those regions (see also Section~S6 in SI), indicating that the PBEsol0+SOC approach with $\alpha=0.25$ is a suitable choice for both materials.

\begin{figure*}[!ht]
\centering
\includegraphics[clip=true,trim=0.7in 5.4in 0.7in 0.7in,scale=0.925]{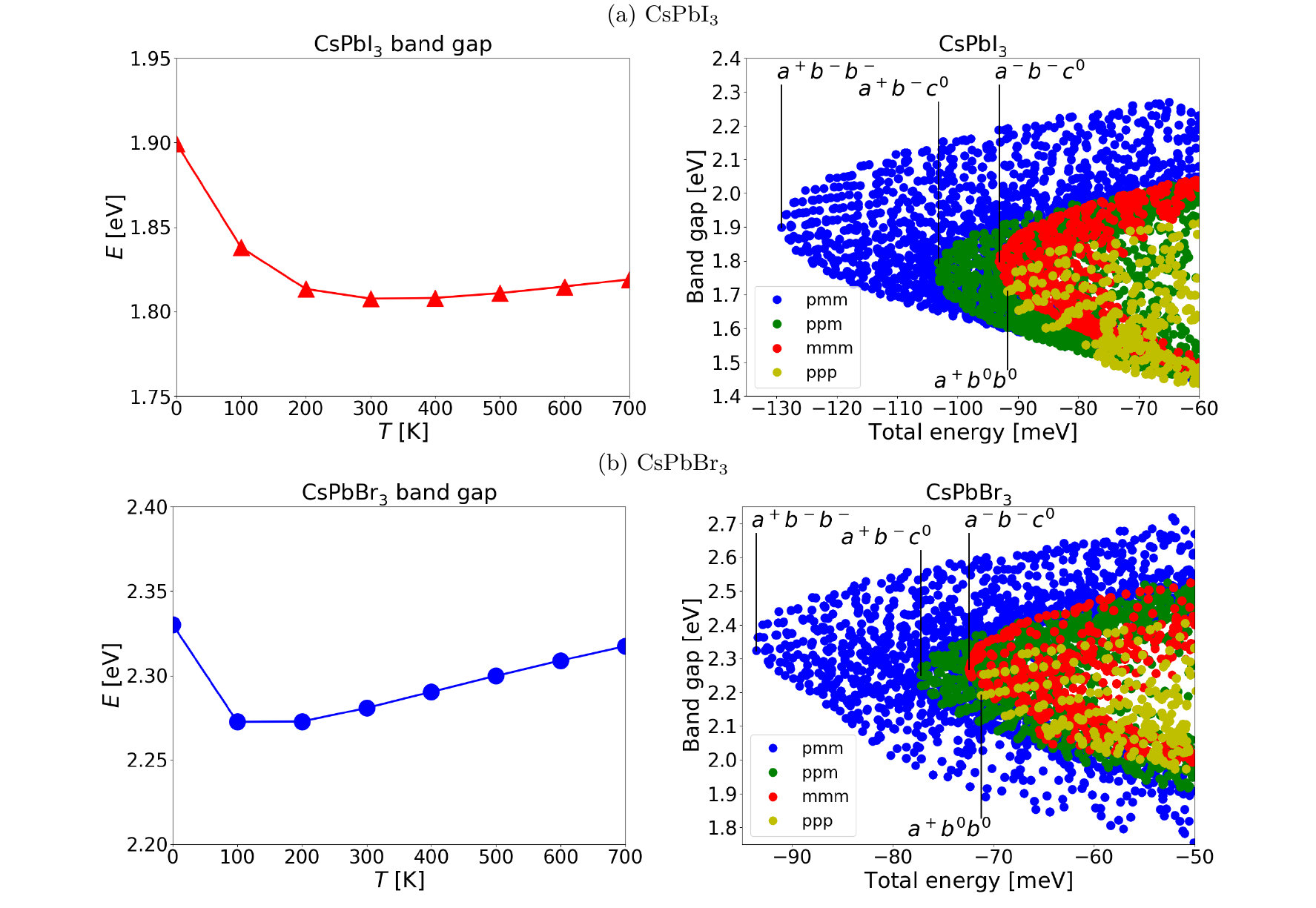}
\caption{Temperature-dependence of the average band gaps (left) and the band-gap vs. total-energy relationship (right) of (a) $\text{CsPbI}_3^{}$ and (b) $\text{CsPbBr}_3^{}\,$. In the right plots, only the BOSS surrogate-model data within the low-energy range are shown. Data from pmm, ppm, mmm, and ppp are colored in blue, green, red, and yellow, respectively. The tilting pattern of the data entry corresponding to the local minimum of each tilting-pattern category is highlighted.}
\label{fig:gaps}
\end{figure*}

Based on the band gap surrogate model, we calculated the temperature-dependent average band gap of each system by
\begin{align*}
\langle E_{\text{gap}}^{}(T) \rangle &=
\frac{\int \rho(\bm{\theta},T) E_{\text{gap}}^{}(\bm{\theta}) \operatorname{d}\bm{\theta}}{\int \rho(\bm{\theta},T) \operatorname{d}\bm{\theta}},
\end{align*}
where the integrations run over the whole configuration space, $E_{\text{gap}}^{}(\bm{\theta})$ is the BOSS-predicted band gap at $\bm{\theta}$, the weighting factor $\rho(\bm{\theta},T)$ is the distribution density displayed in Figure~\ref{fig:thermal}. Results are illustrated in the left column of \textbf{Figure~\ref{fig:gaps}}. The right column of Figure~\ref{fig:gaps} displays the search space-resolved band gap vs. total energy relationship based on the BOSS surrogate model data.

At $0~\text{K}$, both systems are in their $a^+b^-b^-$ global-minimum structures (with some small-amplitude zero-point vibrations). The calculated band gaps are accordingly $1.90$ and $2.33~\text{eV}$ for $\text{CsPbI}_3^{}$ and $\text{CsPbBr}_3^{}\,$, respectively. As temperature increases, structural disorder grows first mainly within pmm (blue in Figure~\ref{fig:gaps}-right). The average band gap decreases as the contribution of ppm (green) increases, since the band gaps of ppm structures within the low-energy range are generally smaller than pmm. A further temperature increase enlarges the contribution from both mmm and ppp to the average band gap. The band gap range of mmm (red) is above ppp (yellow). For $\text{CsPbI}_3^{}$ (Figure~\ref{fig:gaps}a), the band gap range of the union of mmm and ppp approximately superposes ppm. Consequently, the average band gap stabilizes  between $300$ and $700~\text{K}$ ($1.81$ and $1.82~\text{eV}$, respectively). While for $\text{CsPbBr}_3^{}$ (Figure~\ref{fig:gaps}b) at $T>300~\text{K}$, the band gap range of mmm$\oplus$ppp is slightly higher than ppm, thus resulting in a slowly increasing average band gap with temperature ($2.28$ and $2.32~\text{eV}$ at $300$ and $700~\text{K}$, respectively).

For $\text{CsPbBr}_3^{}\,$, the calculated average band gap at room temperature or above agrees well with experiments (ranging between $2.25~\text{eV}$ \cite{Stoumpos2013b} and $2.39~\text{eV}$ \cite{Mannino2020}). Our data reproduce the experimentally observed trend that the band gap increases with temperature above $300~\text{K}$ \cite{Mannino2020}. For $\text{CsPbI}_3^{}\,$, our model somehow overestimates the band gap (reported experimental band gaps are $1.67~\text{eV}$ for the $\upgamma$ phase \cite{Stoumpos2015} and $1.73~\text{eV}$ for the $\upalpha$ phase \cite{Eperon14,Eperon15}). The accuracy of band gap estimation certainly depends on the computational settings (e.g., the amount of exact exchange in a hybrid functional), for which we need benchmark studies in the future. Based on our full-dimensional PESs, however, we can already propose valuable strategies to design next-generation optoelectronic materials. For example, smaller tilting angles are desired to tune the band gap of photovoltaic material $\text{CsPbI}_3^{}$ to the optimal $\sim1.35~\text{eV}$ according to the Shockley-Queisser theory \cite{Shockley61}. Doping with organic monovalent $\text{A}$-site cations that are larger than $\text{Cs}^+$ would be a solution. While for $\text{CsPbBr}_3^{}$ in green-light emission, we are interested in materials with minimal structural disorder for high monochromaticity. $\text{A}$-site mixing with cations smaller than $\text{Cs}^+$ and $\text{X}$-site mixing with larger anions would be promising options.

\section{Conclusions}\label{concl}

This work offers new computational insight into the structural disorder in benchmark systems $\text{CsPbI}_3^{}$ and $\text{CsPbBr}_3^{}\,$. With relatively few first-principles (DFT) calculations sampled with the BOSS scheme, we have obtained  accurate landscapes for both total energy and band gap, each as a function of octahedral-tilting angles. We could infer how the distribution of crystal structures evolves within the three-dimensional octahedral-tilting configuration space with increasing temperature. We then evaluated the statistical mean of band gap, an important materials property for optoelectronic application, over the disordered structures. Our work demonstrates that novel cross-disciplinary machine-learning computational materials science tools such as BOSS can aid the description of structural disorder phenomena, which is important to make recommendation for rational design of advanced functional materials such as halide perovskites.

\section*{Acknowledgments}

We thank Jari J\"arvi, Joakim L\"ofgren, Jarno Laakso, Azimatu Fangnon, Zhaoxin Wu, and Nan Zhang for the fruitful discussions. We acknowledge the computing resources by Xi'an Jiaotong University's HPC platform, the Hefei Advanced Computing Center, the Aalto Science-IT project, and the CSC-IT Center for Science. This work was supported by the National Natural Science Foundation of China under Grant Nos. 62281330043, 11974268, 21503057, 12111530061, and 51902244, the Fundamental Research Funds for the Central Universities (Grant Nos. HIT.NSRIF.2017032 and xzy012021025), the China Postdoctoral Science Foundation (Grant No. 2018M643632), the Natural Science Foundation of Shaanxi Province of China (Grant No. 2023-YBGY-447, 2019JQ-389), the Natural Sciences \& Engineering Research Council of Canada (NSERC, Grant No. RGPIN-2017-06915), the European Union's Horizon 2020 research and innovation program under Grant Agreement No. 676580 [The Novel Materials Discovery (NOMAD)], and the Academy of Finland (Grant Nos. 316601, 334532 and 305632).


\begin{thebibliography}{95}
\expandafter\ifx\csname natexlab\endcsname\relax\def\natexlab#1{#1}\fi
\expandafter\ifx\csname bibnamefont\endcsname\relax
  \def\bibnamefont#1{#1}\fi
\expandafter\ifx\csname bibfnamefont\endcsname\relax
  \def\bibfnamefont#1{#1}\fi
\expandafter\ifx\csname citenamefont\endcsname\relax
  \def\citenamefont#1{#1}\fi
\expandafter\ifx\csname url\endcsname\relax
  \def\url#1{\texttt{#1}}\fi
\expandafter\ifx\csname urlprefix\endcsname\relax\def\urlprefix{URL }\fi
\providecommand{\bibinfo}[2]{#2}
\providecommand{\eprint}[2][]{\url{#2}}

\bibitem[{\citenamefont{Nayak et~al.}(2012)\citenamefont{Nayak,
  Garcia-Belmonte, Kahn, Bisquert, and Cahen}}]{Nayak2012}
\bibinfo{author}{\bibfnamefont{P.~K.} \bibnamefont{Nayak}},
  \bibinfo{author}{\bibfnamefont{G.}~\bibnamefont{Garcia-Belmonte}},
  \bibinfo{author}{\bibfnamefont{A.}~\bibnamefont{Kahn}},
  \bibinfo{author}{\bibfnamefont{J.}~\bibnamefont{Bisquert}}, \bibnamefont{and}
  \bibinfo{author}{\bibfnamefont{D.}~\bibnamefont{Cahen}},
  \bibinfo{journal}{Energy Environ. Sci.} \textbf{\bibinfo{volume}{5}},
  \bibinfo{pages}{6022} (\bibinfo{year}{2012}).

\bibitem[{\citenamefont{Cairns and Goodwin}(2013)}]{Cairns2013}
\bibinfo{author}{\bibfnamefont{A.~B.} \bibnamefont{Cairns}} \bibnamefont{and}
  \bibinfo{author}{\bibfnamefont{A.~L.} \bibnamefont{Goodwin}},
  \bibinfo{journal}{Chem. Soc. Rev.} \textbf{\bibinfo{volume}{42}},
  \bibinfo{pages}{4881} (\bibinfo{year}{2013}).

\bibitem[{\citenamefont{Keen and Goodwin}(2015)}]{Keen2015}
\bibinfo{author}{\bibfnamefont{D.~A.} \bibnamefont{Keen}} \bibnamefont{and}
  \bibinfo{author}{\bibfnamefont{A.~L.} \bibnamefont{Goodwin}},
  \bibinfo{journal}{Nature} \textbf{\bibinfo{volume}{521}},
  \bibinfo{pages}{303} (\bibinfo{year}{2015}).

\bibitem[{\citenamefont{Rhodes et~al.}(2019)\citenamefont{Rhodes, Chae,
  Ribeiro-Palau, and Hone}}]{Rhodes2019}
\bibinfo{author}{\bibfnamefont{D.}~\bibnamefont{Rhodes}},
  \bibinfo{author}{\bibfnamefont{S.~H.} \bibnamefont{Chae}},
  \bibinfo{author}{\bibfnamefont{R.}~\bibnamefont{Ribeiro-Palau}},
  \bibnamefont{and} \bibinfo{author}{\bibfnamefont{J.}~\bibnamefont{Hone}},
  \bibinfo{journal}{Nat. Mater.} \textbf{\bibinfo{volume}{18}},
  \bibinfo{pages}{541} (\bibinfo{year}{2019}).

\bibitem[{\citenamefont{Dragoe and B\'erardan}(2019)}]{Dragoe2019}
\bibinfo{author}{\bibfnamefont{N.}~\bibnamefont{Dragoe}} \bibnamefont{and}
  \bibinfo{author}{\bibfnamefont{D.}~\bibnamefont{B\'erardan}},
  \bibinfo{journal}{Science} \textbf{\bibinfo{volume}{366}},
  \bibinfo{pages}{573} (\bibinfo{year}{2019}).

\bibitem[{\citenamefont{Deringer}(2020)}]{Deringer2020}
\bibinfo{author}{\bibfnamefont{V.~L.} \bibnamefont{Deringer}},
  \bibinfo{journal}{J. Phys. Energy} \textbf{\bibinfo{volume}{2}},
  \bibinfo{pages}{041003} (\bibinfo{year}{2020}).

\bibitem[{\citenamefont{Simonov and Goodwin}(2020)}]{Simonov2020}
\bibinfo{author}{\bibfnamefont{A.}~\bibnamefont{Simonov}} \bibnamefont{and}
  \bibinfo{author}{\bibfnamefont{A.~L.} \bibnamefont{Goodwin}},
  \bibinfo{journal}{Nat. Rev. Chem.} \textbf{\bibinfo{volume}{4}},
  \bibinfo{pages}{657} (\bibinfo{year}{2020}).

\bibitem[{\citenamefont{Deringer et~al.}(2021)\citenamefont{Deringer,
  Bernstein, Cs\'anyi, {Ben~Mahmoud}, Ceriotti, Wilson, Drabold, and
  Elliott}}]{Deringer2021}
\bibinfo{author}{\bibfnamefont{V.~L.} \bibnamefont{Deringer}},
  \bibinfo{author}{\bibfnamefont{N.}~\bibnamefont{Bernstein}},
  \bibinfo{author}{\bibfnamefont{G.}~\bibnamefont{Cs\'anyi}},
  \bibinfo{author}{\bibfnamefont{C.}~\bibnamefont{{Ben~Mahmoud}}},
  \bibinfo{author}{\bibfnamefont{M.}~\bibnamefont{Ceriotti}},
  \bibinfo{author}{\bibfnamefont{M.}~\bibnamefont{Wilson}},
  \bibinfo{author}{\bibfnamefont{D.~A.} \bibnamefont{Drabold}},
  \bibnamefont{and} \bibinfo{author}{\bibfnamefont{S.~R.}
  \bibnamefont{Elliott}}, \bibinfo{journal}{Nature}
  \textbf{\bibinfo{volume}{589}}, \bibinfo{pages}{59} (\bibinfo{year}{2021}).

\bibitem[{\citenamefont{Ye}(2009)}]{YeZG2009}
\bibinfo{author}{\bibfnamefont{Z.-G.} \bibnamefont{Ye}}, \bibinfo{journal}{MRS
  Bullet.} \textbf{\bibinfo{volume}{34}}, \bibinfo{pages}{277}
  (\bibinfo{year}{2009}).

\bibitem[{\citenamefont{Zhang et~al.}(2015)\citenamefont{Zhang, Li, Jiang, Kim,
  Luo, and Geng}}]{ZhangS2015}
\bibinfo{author}{\bibfnamefont{S.}~\bibnamefont{Zhang}},
  \bibinfo{author}{\bibfnamefont{F.}~\bibnamefont{Li}},
  \bibinfo{author}{\bibfnamefont{X.}~\bibnamefont{Jiang}},
  \bibinfo{author}{\bibfnamefont{J.}~\bibnamefont{Kim}},
  \bibinfo{author}{\bibfnamefont{J.}~\bibnamefont{Luo}}, \bibnamefont{and}
  \bibinfo{author}{\bibfnamefont{X.}~\bibnamefont{Geng}},
  \bibinfo{journal}{Prog. Mater. Sci.} \textbf{\bibinfo{volume}{68}},
  \bibinfo{pages}{1} (\bibinfo{year}{2015}).

\bibitem[{\citenamefont{Sun and Cao}(2014)}]{SunE2014}
\bibinfo{author}{\bibfnamefont{E.}~\bibnamefont{Sun}} \bibnamefont{and}
  \bibinfo{author}{\bibfnamefont{W.}~\bibnamefont{Cao}},
  \bibinfo{journal}{Prog. Mater. Sci.} \textbf{\bibinfo{volume}{65}},
  \bibinfo{pages}{124} (\bibinfo{year}{2014}).

\bibitem[{\citenamefont{Abdi-Jalebi et~al.}(2018)\citenamefont{Abdi-Jalebi,
  Andaji-Garmaroudi, Pearson, Divitini, Cacovich, Philippe, Rensmo, Ducati,
  Friend, and Stranks}}]{AbdiJalebi2018a}
\bibinfo{author}{\bibfnamefont{M.}~\bibnamefont{Abdi-Jalebi}},
  \bibinfo{author}{\bibfnamefont{Z.}~\bibnamefont{Andaji-Garmaroudi}},
  \bibinfo{author}{\bibfnamefont{A.~J.} \bibnamefont{Pearson}},
  \bibinfo{author}{\bibfnamefont{G.}~\bibnamefont{Divitini}},
  \bibinfo{author}{\bibfnamefont{S.}~\bibnamefont{Cacovich}},
  \bibinfo{author}{\bibfnamefont{B.}~\bibnamefont{Philippe}},
  \bibinfo{author}{\bibfnamefont{H.}~\bibnamefont{Rensmo}},
  \bibinfo{author}{\bibfnamefont{C.}~\bibnamefont{Ducati}},
  \bibinfo{author}{\bibfnamefont{R.~H.} \bibnamefont{Friend}},
  \bibnamefont{and} \bibinfo{author}{\bibfnamefont{S.~D.}
  \bibnamefont{Stranks}}, \bibinfo{journal}{ACS Energy Lett.}
  \textbf{\bibinfo{volume}{3}}, \bibinfo{pages}{2671} (\bibinfo{year}{2018}).

\bibitem[{\citenamefont{Correa-Baena et~al.}(2019)\citenamefont{Correa-Baena,
  Brenner, Snaider, Sun, Li, Jensen, Hartono, Nienhaus, Wieghold, Poindexter
  et~al.}}]{CorreaBaena2019}
\bibinfo{author}{\bibfnamefont{J.-P.} \bibnamefont{Correa-Baena}},
  \bibinfo{author}{\bibfnamefont{T.~M.} \bibnamefont{Brenner}},
  \bibinfo{author}{\bibfnamefont{J.}~\bibnamefont{Snaider}},
  \bibinfo{author}{\bibfnamefont{S.}~\bibnamefont{Sun}},
  \bibinfo{author}{\bibfnamefont{X.}~\bibnamefont{Li}},
  \bibinfo{author}{\bibfnamefont{M.~A.} \bibnamefont{Jensen}},
  \bibinfo{author}{\bibfnamefont{N.~T.~P.} \bibnamefont{Hartono}},
  \bibinfo{author}{\bibfnamefont{L.}~\bibnamefont{Nienhaus}},
  \bibinfo{author}{\bibfnamefont{S.}~\bibnamefont{Wieghold}},
  \bibinfo{author}{\bibfnamefont{J.~R.} \bibnamefont{Poindexter}},
  \bibnamefont{et~al.}, \bibinfo{journal}{Science}
  \textbf{\bibinfo{volume}{363}}, \bibinfo{pages}{627} (\bibinfo{year}{2019}).

\bibitem[{\citenamefont{Shin et~al.}(2017)\citenamefont{Shin, Saparov, and
  Mitzi}}]{ShinD2017}
\bibinfo{author}{\bibfnamefont{D.}~\bibnamefont{Shin}},
  \bibinfo{author}{\bibfnamefont{B.}~\bibnamefont{Saparov}}, \bibnamefont{and}
  \bibinfo{author}{\bibfnamefont{D.~B.} \bibnamefont{Mitzi}},
  \bibinfo{journal}{Adv. Energy Mater.} \textbf{\bibinfo{volume}{7}},
  \bibinfo{pages}{1602366} (\bibinfo{year}{2017}).

\bibitem[{\citenamefont{Giraldo et~al.}(2019)\citenamefont{Giraldo, Jehl,
  Placidi, Izquierdo-Roca, P\'erez-Rodr\'iguez, and Saucedo}}]{Giraldo2019}
\bibinfo{author}{\bibfnamefont{S.}~\bibnamefont{Giraldo}},
  \bibinfo{author}{\bibfnamefont{Z.}~\bibnamefont{Jehl}},
  \bibinfo{author}{\bibfnamefont{M.}~\bibnamefont{Placidi}},
  \bibinfo{author}{\bibfnamefont{V.}~\bibnamefont{Izquierdo-Roca}},
  \bibinfo{author}{\bibfnamefont{A.}~\bibnamefont{P\'erez-Rodr\'iguez}},
  \bibnamefont{and} \bibinfo{author}{\bibfnamefont{E.}~\bibnamefont{Saucedo}},
  \bibinfo{journal}{Adv. Mater.} \textbf{\bibinfo{volume}{31}},
  \bibinfo{pages}{1806692} (\bibinfo{year}{2019}).

\bibitem[{\citenamefont{Tan et~al.}(2022)\citenamefont{Tan, Yu, Cui, Meng,
  Huang, Li, Chen, Wu, Shi, Luo et~al.}}]{TanS2022}
\bibinfo{author}{\bibfnamefont{S.}~\bibnamefont{Tan}},
  \bibinfo{author}{\bibfnamefont{B.}~\bibnamefont{Yu}},
  \bibinfo{author}{\bibfnamefont{Y.}~\bibnamefont{Cui}},
  \bibinfo{author}{\bibfnamefont{F.}~\bibnamefont{Meng}},
  \bibinfo{author}{\bibfnamefont{C.}~\bibnamefont{Huang}},
  \bibinfo{author}{\bibfnamefont{Y.}~\bibnamefont{Li}},
  \bibinfo{author}{\bibfnamefont{Z.}~\bibnamefont{Chen}},
  \bibinfo{author}{\bibfnamefont{H.}~\bibnamefont{Wu}},
  \bibinfo{author}{\bibfnamefont{J.}~\bibnamefont{Shi}},
  \bibinfo{author}{\bibfnamefont{Y.}~\bibnamefont{Luo}}, \bibnamefont{et~al.},
  \bibinfo{journal}{Angew. Chem. Int. Ed.} \textbf{\bibinfo{volume}{61}},
  \bibinfo{pages}{e202201300} (\bibinfo{year}{2022}).

\bibitem[{\citenamefont{Stoumpos and Kanatzidis}(2015)}]{Stoumpos2015}
\bibinfo{author}{\bibfnamefont{C.~C.} \bibnamefont{Stoumpos}} \bibnamefont{and}
  \bibinfo{author}{\bibfnamefont{M.~G.} \bibnamefont{Kanatzidis}},
  \bibinfo{journal}{Acc. Chem. Res.} \textbf{\bibinfo{volume}{48}},
  \bibinfo{pages}{2791} (\bibinfo{year}{2015}).

\bibitem[{\citenamefont{Marronnier et~al.}(2018)\citenamefont{Marronnier, Roma,
  Boyer-Richard, Pedesseau, Jancu, Bonnassieux, Katan, Stoumpos, Kanatzidis,
  and Even}}]{Marronnier2018}
\bibinfo{author}{\bibfnamefont{A.}~\bibnamefont{Marronnier}},
  \bibinfo{author}{\bibfnamefont{G.}~\bibnamefont{Roma}},
  \bibinfo{author}{\bibfnamefont{S.}~\bibnamefont{Boyer-Richard}},
  \bibinfo{author}{\bibfnamefont{L.}~\bibnamefont{Pedesseau}},
  \bibinfo{author}{\bibfnamefont{J.-M.} \bibnamefont{Jancu}},
  \bibinfo{author}{\bibfnamefont{Y.}~\bibnamefont{Bonnassieux}},
  \bibinfo{author}{\bibfnamefont{C.}~\bibnamefont{Katan}},
  \bibinfo{author}{\bibfnamefont{C.~C.} \bibnamefont{Stoumpos}},
  \bibinfo{author}{\bibfnamefont{M.~G.} \bibnamefont{Kanatzidis}},
  \bibnamefont{and} \bibinfo{author}{\bibfnamefont{J.}~\bibnamefont{Even}},
  \bibinfo{journal}{ACS Nano} \textbf{\bibinfo{volume}{12}},
  \bibinfo{pages}{3477} (\bibinfo{year}{2018}).

\bibitem[{\citenamefont{Jinnouchi et~al.}(2019)\citenamefont{Jinnouchi,
  Lahnsteiner, Karsai, Kresse, and Bokdam}}]{Jinnouchi2019}
\bibinfo{author}{\bibfnamefont{R.}~\bibnamefont{Jinnouchi}},
  \bibinfo{author}{\bibfnamefont{J.}~\bibnamefont{Lahnsteiner}},
  \bibinfo{author}{\bibfnamefont{F.}~\bibnamefont{Karsai}},
  \bibinfo{author}{\bibfnamefont{G.}~\bibnamefont{Kresse}}, \bibnamefont{and}
  \bibinfo{author}{\bibfnamefont{M.}~\bibnamefont{Bokdam}},
  \bibinfo{journal}{Phys. Rev. Lett.} \textbf{\bibinfo{volume}{122}},
  \bibinfo{pages}{225701} (\bibinfo{year}{2019}).

\bibitem[{\citenamefont{Klarbring}(2019)}]{Klarbring2019}
\bibinfo{author}{\bibfnamefont{J.}~\bibnamefont{Klarbring}},
  \bibinfo{journal}{Phys. Rev. B} \textbf{\bibinfo{volume}{949}},
  \bibinfo{pages}{104105} (\bibinfo{year}{2019}).

\bibitem[{\citenamefont{Chen et~al.}(2020)\citenamefont{Chen, Xu, Yang, and
  Bellaiche}}]{ChenL2020}
\bibinfo{author}{\bibfnamefont{L.}~\bibnamefont{Chen}},
  \bibinfo{author}{\bibfnamefont{B.}~\bibnamefont{Xu}},
  \bibinfo{author}{\bibfnamefont{Y.}~\bibnamefont{Yang}}, \bibnamefont{and}
  \bibinfo{author}{\bibfnamefont{L.}~\bibnamefont{Bellaiche}},
  \bibinfo{journal}{Adv. Funct. Mater.} \textbf{\bibinfo{volume}{30}},
  \bibinfo{pages}{1909496} (\bibinfo{year}{2020}).

\bibitem[{\citenamefont{Patrick et~al.}(2015)\citenamefont{Patrick, Jacobsen,
  and Thygesen}}]{Patrick15}
\bibinfo{author}{\bibfnamefont{C.~E.} \bibnamefont{Patrick}},
  \bibinfo{author}{\bibfnamefont{K.~W.} \bibnamefont{Jacobsen}},
  \bibnamefont{and} \bibinfo{author}{\bibfnamefont{K.~S.}
  \bibnamefont{Thygesen}}, \bibinfo{journal}{Phys. Rev. B}
  \textbf{\bibinfo{volume}{92}}, \bibinfo{pages}{201205}
  (\bibinfo{year}{2015}).

\bibitem[{\citenamefont{Yang et~al.}(2020)\citenamefont{Yang, Skelton,
  {da~Silva}, Frost, and Walsh}}]{YangRX2020}
\bibinfo{author}{\bibfnamefont{R.~X.} \bibnamefont{Yang}},
  \bibinfo{author}{\bibfnamefont{J.~M.} \bibnamefont{Skelton}},
  \bibinfo{author}{\bibfnamefont{E.~L.} \bibnamefont{{da~Silva}}},
  \bibinfo{author}{\bibfnamefont{J.~M.} \bibnamefont{Frost}}, \bibnamefont{and}
  \bibinfo{author}{\bibfnamefont{A.}~\bibnamefont{Walsh}}, \bibinfo{journal}{J.
  Phys. Chem. Lett.} \textbf{\bibinfo{volume}{152}}, \bibinfo{pages}{024703}
  (\bibinfo{year}{2020}).

\bibitem[{\citenamefont{Zhu et~al.}(2022)\citenamefont{Zhu, {Caicedo-D\'avila},
  Gehrmann, and Egger}}]{ZhuX2022}
\bibinfo{author}{\bibfnamefont{X.}~\bibnamefont{Zhu}},
  \bibinfo{author}{\bibfnamefont{S.}~\bibnamefont{{Caicedo-D\'avila}}},
  \bibinfo{author}{\bibfnamefont{C.}~\bibnamefont{Gehrmann}}, \bibnamefont{and}
  \bibinfo{author}{\bibfnamefont{D.~A.} \bibnamefont{Egger}},
  \bibinfo{journal}{ACS Appl. Mater. Interfaces} \textbf{\bibinfo{volume}{14}},
  \bibinfo{pages}{22973} (\bibinfo{year}{2022}).

\bibitem[{\citenamefont{Kirkpatrick et~al.}(1983)\citenamefont{Kirkpatrick,
  Gellatt, and Vecchi}}]{Kirkpatrick1983}
\bibinfo{author}{\bibfnamefont{S.}~\bibnamefont{Kirkpatrick}},
  \bibinfo{author}{\bibfnamefont{C.~G.} \bibnamefont{Gellatt}},
  \bibnamefont{and} \bibinfo{author}{\bibfnamefont{M.~P.}
  \bibnamefont{Vecchi}}, \bibinfo{journal}{Science}
  \textbf{\bibinfo{volume}{220}}, \bibinfo{pages}{671} (\bibinfo{year}{1983}).

\bibitem[{\citenamefont{Goedecker et~al.}(2005)\citenamefont{Goedecker,
  Hellmann, and Lenosky}}]{Goedecker2005}
\bibinfo{author}{\bibfnamefont{S.}~\bibnamefont{Goedecker}},
  \bibinfo{author}{\bibfnamefont{W.}~\bibnamefont{Hellmann}}, \bibnamefont{and}
  \bibinfo{author}{\bibfnamefont{T.}~\bibnamefont{Lenosky}},
  \bibinfo{journal}{Phys. Rev. Lett.} \textbf{\bibinfo{volume}{95}},
  \bibinfo{pages}{055501} (\bibinfo{year}{2005}).

\bibitem[{\citenamefont{Jones}(2015)}]{JonesRO2015}
\bibinfo{author}{\bibfnamefont{R.~O.} \bibnamefont{Jones}},
  \bibinfo{journal}{Rev. Mod. Phys.} \textbf{\bibinfo{volume}{87}},
  \bibinfo{pages}{897} (\bibinfo{year}{2015}).

\bibitem[{\citenamefont{Bisseling et~al.}(1986)\citenamefont{Bisseling,
  Kosloff, and Kosloff}}]{Bisseling1986}
\bibinfo{author}{\bibfnamefont{R.~H.} \bibnamefont{Bisseling}},
  \bibinfo{author}{\bibfnamefont{R.}~\bibnamefont{Kosloff}}, \bibnamefont{and}
  \bibinfo{author}{\bibfnamefont{D.}~\bibnamefont{Kosloff}},
  \bibinfo{journal}{Comput. Phys. Commun.} \textbf{\bibinfo{volume}{39}},
  \bibinfo{pages}{313} (\bibinfo{year}{1986}).

\bibitem[{\citenamefont{Wu et~al.}(2004)\citenamefont{Wu, Werner, and
  Manthe}}]{WuT2004}
\bibinfo{author}{\bibfnamefont{T.}~\bibnamefont{Wu}},
  \bibinfo{author}{\bibfnamefont{H.-J.} \bibnamefont{Werner}},
  \bibnamefont{and} \bibinfo{author}{\bibfnamefont{U.}~\bibnamefont{Manthe}},
  \bibinfo{journal}{Science} \textbf{\bibinfo{volume}{306}},
  \bibinfo{pages}{2227} (\bibinfo{year}{2004}).

\bibitem[{\citenamefont{Schleder et~al.}(2019)\citenamefont{Schleder, Padilha,
  {Mera~Acosta}, Costa, and Fazzio}}]{Schleder2019b}
\bibinfo{author}{\bibfnamefont{G.~R.} \bibnamefont{Schleder}},
  \bibinfo{author}{\bibfnamefont{A.~C.~M.} \bibnamefont{Padilha}},
  \bibinfo{author}{\bibfnamefont{C.}~\bibnamefont{{Mera~Acosta}}},
  \bibinfo{author}{\bibfnamefont{M.}~\bibnamefont{Costa}}, \bibnamefont{and}
  \bibinfo{author}{\bibfnamefont{A.}~\bibnamefont{Fazzio}},
  \bibinfo{journal}{J. Phys.: Mater.} \textbf{\bibinfo{volume}{2}},
  \bibinfo{pages}{032001} (\bibinfo{year}{2019}).

\bibitem[{\citenamefont{Schmidt et~al.}(2019)\citenamefont{Schmidt, Marques,
  Botti, and Marques}}]{Schmidt2019}
\bibinfo{author}{\bibfnamefont{J.}~\bibnamefont{Schmidt}},
  \bibinfo{author}{\bibfnamefont{M.~R.~G.} \bibnamefont{Marques}},
  \bibinfo{author}{\bibfnamefont{S.}~\bibnamefont{Botti}}, \bibnamefont{and}
  \bibinfo{author}{\bibfnamefont{M.~A.~L.} \bibnamefont{Marques}},
  \bibinfo{journal}{npj Comput. Mater.} \textbf{\bibinfo{volume}{5}},
  \bibinfo{pages}{83} (\bibinfo{year}{2019}).

\bibitem[{\citenamefont{Himanen et~al.}(2019)\citenamefont{Himanen, Geurts,
  Foster, and Rinke}}]{Himanen2019}
\bibinfo{author}{\bibfnamefont{L.}~\bibnamefont{Himanen}},
  \bibinfo{author}{\bibfnamefont{A.}~\bibnamefont{Geurts}},
  \bibinfo{author}{\bibfnamefont{A.~S.} \bibnamefont{Foster}},
  \bibnamefont{and} \bibinfo{author}{\bibfnamefont{P.}~\bibnamefont{Rinke}},
  \bibinfo{journal}{Adv. Sci.} \textbf{\bibinfo{volume}{6}},
  \bibinfo{pages}{1900808} (\bibinfo{year}{2019}).

\bibitem[{\citenamefont{Deringer et~al.}(2019)\citenamefont{Deringer, Caro, and
  Cs\'anyi}}]{Deringer2019}
\bibinfo{author}{\bibfnamefont{V.~L.} \bibnamefont{Deringer}},
  \bibinfo{author}{\bibfnamefont{M.~A.} \bibnamefont{Caro}}, \bibnamefont{and}
  \bibinfo{author}{\bibfnamefont{G.}~\bibnamefont{Cs\'anyi}},
  \bibinfo{journal}{Adv. Mater.} \textbf{\bibinfo{volume}{31}},
  \bibinfo{pages}{1902765} (\bibinfo{year}{2019}).

\bibitem[{\citenamefont{Hart et~al.}(2021)\citenamefont{Hart, Mueller, Toher,
  and Curtarolo}}]{Hart2021}
\bibinfo{author}{\bibfnamefont{G.~L.~W.} \bibnamefont{Hart}},
  \bibinfo{author}{\bibfnamefont{T.}~\bibnamefont{Mueller}},
  \bibinfo{author}{\bibfnamefont{C.}~\bibnamefont{Toher}}, \bibnamefont{and}
  \bibinfo{author}{\bibfnamefont{S.}~\bibnamefont{Curtarolo}},
  \bibinfo{journal}{Nat. Rev. Mater.} \textbf{\bibinfo{volume}{6}},
  \bibinfo{pages}{730} (\bibinfo{year}{2021}).

\bibitem[{\citenamefont{Bart\'ok et~al.}(2010)\citenamefont{Bart\'ok, Payne,
  Kondor, and Cs\'anyi}}]{Bartok2010}
\bibinfo{author}{\bibfnamefont{A.~P.} \bibnamefont{Bart\'ok}},
  \bibinfo{author}{\bibfnamefont{M.~C.} \bibnamefont{Payne}},
  \bibinfo{author}{\bibfnamefont{R.}~\bibnamefont{Kondor}}, \bibnamefont{and}
  \bibinfo{author}{\bibfnamefont{G.}~\bibnamefont{Cs\'anyi}},
  \bibinfo{journal}{Phys. Rev. Lett.} \textbf{\bibinfo{volume}{104}},
  \bibinfo{pages}{136403} (\bibinfo{year}{2010}).

\bibitem[{\citenamefont{Behler}(2014)}]{Behler2014}
\bibinfo{author}{\bibfnamefont{J.}~\bibnamefont{Behler}}, \bibinfo{journal}{J.
  Phys.: Condens. Matter} \textbf{\bibinfo{volume}{26}},
  \bibinfo{pages}{183001} (\bibinfo{year}{2014}).

\bibitem[{\citenamefont{d'Avezac and Zunger}(2008)}]{dAvezac2008}
\bibinfo{author}{\bibfnamefont{M.}~\bibnamefont{d'Avezac}} \bibnamefont{and}
  \bibinfo{author}{\bibfnamefont{A.}~\bibnamefont{Zunger}},
  \bibinfo{journal}{Phys. Rev. B} \textbf{\bibinfo{volume}{78}},
  \bibinfo{pages}{064102} (\bibinfo{year}{2008}).

\bibitem[{\citenamefont{Wang et~al.}(2012)\citenamefont{Wang, Lv, Zhu, and
  Ma}}]{WangY2012}
\bibinfo{author}{\bibfnamefont{Y.}~\bibnamefont{Wang}},
  \bibinfo{author}{\bibfnamefont{J.}~\bibnamefont{Lv}},
  \bibinfo{author}{\bibfnamefont{L.}~\bibnamefont{Zhu}}, \bibnamefont{and}
  \bibinfo{author}{\bibfnamefont{Y.}~\bibnamefont{Ma}},
  \bibinfo{journal}{Comput. Phys. Commun.} \textbf{\bibinfo{volume}{183}},
  \bibinfo{pages}{2063} (\bibinfo{year}{2012}).

\bibitem[{\citenamefont{Bhattacharya et~al.}(2013)\citenamefont{Bhattacharya,
  Levchenko, Ghiringhelli, and Scheffler}}]{Bhattacharya2013}
\bibinfo{author}{\bibfnamefont{S.}~\bibnamefont{Bhattacharya}},
  \bibinfo{author}{\bibfnamefont{S.~V.} \bibnamefont{Levchenko}},
  \bibinfo{author}{\bibfnamefont{L.~M.} \bibnamefont{Ghiringhelli}},
  \bibnamefont{and}
  \bibinfo{author}{\bibfnamefont{M.}~\bibnamefont{Scheffler}},
  \bibinfo{journal}{Phys. Rev. Lett.} \textbf{\bibinfo{volume}{111}},
  \bibinfo{pages}{135501} (\bibinfo{year}{2013}).

\bibitem[{\citenamefont{Yamashita et~al.}(2018)\citenamefont{Yamashita, Sato,
  Kino, Miyake, Tsuda, and Oguchi}}]{Yamashita2018}
\bibinfo{author}{\bibfnamefont{T.}~\bibnamefont{Yamashita}},
  \bibinfo{author}{\bibfnamefont{N.}~\bibnamefont{Sato}},
  \bibinfo{author}{\bibfnamefont{H.}~\bibnamefont{Kino}},
  \bibinfo{author}{\bibfnamefont{T.}~\bibnamefont{Miyake}},
  \bibinfo{author}{\bibfnamefont{K.}~\bibnamefont{Tsuda}}, \bibnamefont{and}
  \bibinfo{author}{\bibfnamefont{T.}~\bibnamefont{Oguchi}},
  \bibinfo{journal}{Phys. Rev. Materials} \textbf{\bibinfo{volume}{2}},
  \bibinfo{pages}{013803} (\bibinfo{year}{2018}).

\bibitem[{\citenamefont{J\o{}rgensen et~al.}(2018)\citenamefont{J\o{}rgensen,
  Larsen, Jacobsen, and Hammer}}]{Joergensen2018}
\bibinfo{author}{\bibfnamefont{M.~S.} \bibnamefont{J\o{}rgensen}},
  \bibinfo{author}{\bibfnamefont{U.~F.} \bibnamefont{Larsen}},
  \bibinfo{author}{\bibfnamefont{K.~W.} \bibnamefont{Jacobsen}},
  \bibnamefont{and} \bibinfo{author}{\bibfnamefont{B.}~\bibnamefont{Hammer}},
  \bibinfo{journal}{J. Phys. Chem. A} \textbf{\bibinfo{volume}{122}},
  \bibinfo{pages}{1504} (\bibinfo{year}{2018}).

\bibitem[{\citenamefont{Bisbo and Hammer}(2020)}]{Bisbo2020}
\bibinfo{author}{\bibfnamefont{M.~K.} \bibnamefont{Bisbo}} \bibnamefont{and}
  \bibinfo{author}{\bibfnamefont{B.}~\bibnamefont{Hammer}},
  \bibinfo{journal}{Phys. Rev. Lett.} \textbf{\bibinfo{volume}{124}},
  \bibinfo{pages}{086102} (\bibinfo{year}{2020}).

\bibitem[{\citenamefont{Mortensen et~al.}(2020)\citenamefont{Mortensen,
  Meldgaard, Bisbo, Christiansen, and Hammer}}]{Mortensen2020}
\bibinfo{author}{\bibfnamefont{H.~L.} \bibnamefont{Mortensen}},
  \bibinfo{author}{\bibfnamefont{S.~A.} \bibnamefont{Meldgaard}},
  \bibinfo{author}{\bibfnamefont{M.~K.} \bibnamefont{Bisbo}},
  \bibinfo{author}{\bibfnamefont{M.-P.~V.} \bibnamefont{Christiansen}},
  \bibnamefont{and} \bibinfo{author}{\bibfnamefont{B.}~\bibnamefont{Hammer}},
  \bibinfo{journal}{Phys. Rev. B} \textbf{\bibinfo{volume}{102}},
  \bibinfo{pages}{075427} (\bibinfo{year}{2020}).

\bibitem[{\citenamefont{Kaappa et~al.}(2021)\citenamefont{Kaappa,
  {del~R\'{\i}o}, and Jacobsen}}]{Kaappa2021}
\bibinfo{author}{\bibfnamefont{S.}~\bibnamefont{Kaappa}},
  \bibinfo{author}{\bibfnamefont{E.~G.} \bibnamefont{{del~R\'{\i}o}}},
  \bibnamefont{and} \bibinfo{author}{\bibfnamefont{K.~W.}
  \bibnamefont{Jacobsen}}, \bibinfo{journal}{Phys. Rev. B}
  \textbf{\bibinfo{volume}{103}}, \bibinfo{pages}{174114}
  (\bibinfo{year}{2021}).

\bibitem[{\citenamefont{Wanzenb\"ock et~al.}(2022)\citenamefont{Wanzenb\"ock,
  Marco, Sebastian, Florian, Carrete, and Madsen}}]{Wanzenboeck2022}
\bibinfo{author}{\bibfnamefont{R.}~\bibnamefont{Wanzenb\"ock}},
  \bibinfo{author}{\bibfnamefont{A.}~\bibnamefont{Marco}},
  \bibinfo{author}{\bibfnamefont{B.}~\bibnamefont{Sebastian}},
  \bibinfo{author}{\bibfnamefont{B.}~\bibnamefont{Florian}},
  \bibinfo{author}{\bibfnamefont{J.}~\bibnamefont{Carrete}}, \bibnamefont{and}
  \bibinfo{author}{\bibfnamefont{G.~K.~H.} \bibnamefont{Madsen}},
  \bibinfo{journal}{Digital Discovery} \textbf{\bibinfo{volume}{1}},
  \bibinfo{pages}{703} (\bibinfo{year}{2022}).

\bibitem[{\citenamefont{Rupp et~al.}(2012)\citenamefont{Rupp, Tkatchenko,
  M\"uller, and {von~Lilienfeld}}}]{Rupp2012}
\bibinfo{author}{\bibfnamefont{M.}~\bibnamefont{Rupp}},
  \bibinfo{author}{\bibfnamefont{A.}~\bibnamefont{Tkatchenko}},
  \bibinfo{author}{\bibfnamefont{K.-R.} \bibnamefont{M\"uller}},
  \bibnamefont{and} \bibinfo{author}{\bibfnamefont{O.~A.}
  \bibnamefont{{von~Lilienfeld}}}, \bibinfo{journal}{Phys. Rev. Lett.}
  \textbf{\bibinfo{volume}{108}}, \bibinfo{pages}{058301}
  (\bibinfo{year}{2012}).

\bibitem[{\citenamefont{Botu et~al.}(2017)\citenamefont{Botu, Batra, Chapman,
  and Ramprasad}}]{Botu2017}
\bibinfo{author}{\bibfnamefont{V.}~\bibnamefont{Botu}},
  \bibinfo{author}{\bibfnamefont{R.}~\bibnamefont{Batra}},
  \bibinfo{author}{\bibfnamefont{J.}~\bibnamefont{Chapman}}, \bibnamefont{and}
  \bibinfo{author}{\bibfnamefont{R.}~\bibnamefont{Ramprasad}},
  \bibinfo{journal}{J. Phys. Chem. C} \textbf{\bibinfo{volume}{121}},
  \bibinfo{pages}{511} (\bibinfo{year}{2017}).

\bibitem[{\citenamefont{{von~Lilienfeld} and Burke}(2020)}]{vonLilienfeld2020}
\bibinfo{author}{\bibfnamefont{O.~A.} \bibnamefont{{von~Lilienfeld}}}
  \bibnamefont{and} \bibinfo{author}{\bibfnamefont{K.}~\bibnamefont{Burke}},
  \bibinfo{journal}{Nature Commun.} \textbf{\bibinfo{volume}{11}},
  \bibinfo{pages}{4895} (\bibinfo{year}{2020}).

\bibitem[{\citenamefont{Sauceda et~al.}(2020)\citenamefont{Sauceda, Chmiela,
  Poltavsky, M{\"u}ller, and Tkatchenko}}]{Sauceda2020}
\bibinfo{author}{\bibfnamefont{H.~E.} \bibnamefont{Sauceda}},
  \bibinfo{author}{\bibfnamefont{S.}~\bibnamefont{Chmiela}},
  \bibinfo{author}{\bibfnamefont{I.}~\bibnamefont{Poltavsky}},
  \bibinfo{author}{\bibfnamefont{K.-R.} \bibnamefont{M{\"u}ller}},
  \bibnamefont{and}
  \bibinfo{author}{\bibfnamefont{A.}~\bibnamefont{Tkatchenko}}, in
  \emph{\bibinfo{booktitle}{Machine Learning Meets Quantum Physics}}, edited by
  \bibinfo{editor}{\bibfnamefont{K.~T.} \bibnamefont{Sch{\"u}tt}},
  \bibinfo{editor}{\bibfnamefont{S.}~\bibnamefont{Chmiela}},
  \bibinfo{editor}{\bibfnamefont{O.~A.} \bibnamefont{{von Lilienfeld}}},
  \bibinfo{editor}{\bibfnamefont{A.}~\bibnamefont{Tkatchenko}},
  \bibinfo{editor}{\bibfnamefont{K.}~\bibnamefont{Tsuda}}, \bibnamefont{and}
  \bibinfo{editor}{\bibfnamefont{K.-R.} \bibnamefont{M{\"u}ller}}
  (\bibinfo{publisher}{Springer International Publishing},
  \bibinfo{address}{Cham}, \bibinfo{year}{2020}), p. \bibinfo{pages}{277}.

\bibitem[{\citenamefont{Vandermause et~al.}(2020)\citenamefont{Vandermause,
  Torrisi, Batzner, Xie, Sun, Kolpak, and Kozinsky}}]{Vandermause2020}
\bibinfo{author}{\bibfnamefont{J.}~\bibnamefont{Vandermause}},
  \bibinfo{author}{\bibfnamefont{S.~B.} \bibnamefont{Torrisi}},
  \bibinfo{author}{\bibfnamefont{S.}~\bibnamefont{Batzner}},
  \bibinfo{author}{\bibfnamefont{Y.}~\bibnamefont{Xie}},
  \bibinfo{author}{\bibfnamefont{L.}~\bibnamefont{Sun}},
  \bibinfo{author}{\bibfnamefont{A.~M.} \bibnamefont{Kolpak}},
  \bibnamefont{and} \bibinfo{author}{\bibfnamefont{B.}~\bibnamefont{Kozinsky}},
  \bibinfo{journal}{npj Comput. Mater.} p.~\bibinfo{pages}{20}
  (\bibinfo{year}{2020}).

\bibitem[{\citenamefont{Timmermann et~al.}(2021)\citenamefont{Timmermann, Lee,
  Staacke, Margraf, Scheurer, and Reuter}}]{Timmermann2021}
\bibinfo{author}{\bibfnamefont{J.}~\bibnamefont{Timmermann}},
  \bibinfo{author}{\bibfnamefont{Y.}~\bibnamefont{Lee}},
  \bibinfo{author}{\bibfnamefont{C.~G.} \bibnamefont{Staacke}},
  \bibinfo{author}{\bibfnamefont{J.~T.} \bibnamefont{Margraf}},
  \bibinfo{author}{\bibfnamefont{C.}~\bibnamefont{Scheurer}}, \bibnamefont{and}
  \bibinfo{author}{\bibfnamefont{K.}~\bibnamefont{Reuter}},
  \bibinfo{journal}{J. Chem. Phys.} \textbf{\bibinfo{volume}{155}},
  \bibinfo{pages}{244107} (\bibinfo{year}{2021}).

\bibitem[{\citenamefont{Unke et~al.}(2021)\citenamefont{Unke, Chmiela, Sauceda,
  Gastegger, Poltavsky, Sch\"utt, Tkatchenko, and M\"uller}}]{Unke2021}
\bibinfo{author}{\bibfnamefont{O.~T.} \bibnamefont{Unke}},
  \bibinfo{author}{\bibfnamefont{S.}~\bibnamefont{Chmiela}},
  \bibinfo{author}{\bibfnamefont{H.~E.} \bibnamefont{Sauceda}},
  \bibinfo{author}{\bibfnamefont{M.}~\bibnamefont{Gastegger}},
  \bibinfo{author}{\bibfnamefont{I.}~\bibnamefont{Poltavsky}},
  \bibinfo{author}{\bibfnamefont{K.~T.} \bibnamefont{Sch\"utt}},
  \bibinfo{author}{\bibfnamefont{A.}~\bibnamefont{Tkatchenko}},
  \bibnamefont{and} \bibinfo{author}{\bibfnamefont{K.-R.}
  \bibnamefont{M\"uller}}, \bibinfo{journal}{Chem. Rev.}
  \textbf{\bibinfo{volume}{121}}, \bibinfo{pages}{10142}
  (\bibinfo{year}{2021}).

\bibitem[{\citenamefont{Westermayr et~al.}(2022)\citenamefont{Westermayr,
  Chaudhuri, Jeindl, Hofmann, and Maurer}}]{Westermayr2022}
\bibinfo{author}{\bibfnamefont{J.}~\bibnamefont{Westermayr}},
  \bibinfo{author}{\bibfnamefont{S.}~\bibnamefont{Chaudhuri}},
  \bibinfo{author}{\bibfnamefont{A.}~\bibnamefont{Jeindl}},
  \bibinfo{author}{\bibfnamefont{O.~T.} \bibnamefont{Hofmann}},
  \bibnamefont{and} \bibinfo{author}{\bibfnamefont{R.~J.}
  \bibnamefont{Maurer}}, \bibinfo{journal}{Digital Discovery}
  \textbf{\bibinfo{volume}{1}}, \bibinfo{pages}{463} (\bibinfo{year}{2022}).

\bibitem[{\citenamefont{Todorovi\'c et~al.}(2019)\citenamefont{Todorovi\'c,
  Gutman, Corander, and Rinke}}]{Todorovic2019}
\bibinfo{author}{\bibfnamefont{M.}~\bibnamefont{Todorovi\'c}},
  \bibinfo{author}{\bibfnamefont{M.~U.} \bibnamefont{Gutman}},
  \bibinfo{author}{\bibfnamefont{J.}~\bibnamefont{Corander}}, \bibnamefont{and}
  \bibinfo{author}{\bibfnamefont{P.}~\bibnamefont{Rinke}},
  \bibinfo{journal}{npj Comput. Mater.} \textbf{\bibinfo{volume}{5}},
  \bibinfo{pages}{35} (\bibinfo{year}{2019}).

\bibitem[{\citenamefont{Fang et~al.}(2021)\citenamefont{Fang, Makkonen,
  Todorov\'c, Rinke, and Chen}}]{FangL2021}
\bibinfo{author}{\bibfnamefont{L.}~\bibnamefont{Fang}},
  \bibinfo{author}{\bibfnamefont{E.}~\bibnamefont{Makkonen}},
  \bibinfo{author}{\bibfnamefont{M.}~\bibnamefont{Todorov\'c}},
  \bibinfo{author}{\bibfnamefont{P.}~\bibnamefont{Rinke}}, \bibnamefont{and}
  \bibinfo{author}{\bibfnamefont{X.}~\bibnamefont{Chen}}, \bibinfo{journal}{J.
  Chem. Theory Comput.} \textbf{\bibinfo{volume}{17}}, \bibinfo{pages}{1955}
  (\bibinfo{year}{2021}).

\bibitem[{\citenamefont{Egger et~al.}(2020)\citenamefont{Egger, H\"ormann,
  Jeindl, Scherbela, Obersteiner, Todorovi\'{c}, Rinke, and
  Hofmann}}]{EggerAT2020}
\bibinfo{author}{\bibfnamefont{A.~T.} \bibnamefont{Egger}},
  \bibinfo{author}{\bibfnamefont{L.}~\bibnamefont{H\"ormann}},
  \bibinfo{author}{\bibfnamefont{A.}~\bibnamefont{Jeindl}},
  \bibinfo{author}{\bibfnamefont{M.}~\bibnamefont{Scherbela}},
  \bibinfo{author}{\bibfnamefont{V.}~\bibnamefont{Obersteiner}},
  \bibinfo{author}{\bibfnamefont{M.}~\bibnamefont{Todorovi\'{c}}},
  \bibinfo{author}{\bibfnamefont{P.}~\bibnamefont{Rinke}}, \bibnamefont{and}
  \bibinfo{author}{\bibfnamefont{O.~T.} \bibnamefont{Hofmann}},
  \bibinfo{journal}{Adv. Sci.} \textbf{\bibinfo{volume}{7}},
  \bibinfo{pages}{2000992} (\bibinfo{year}{2020}).

\bibitem[{\citenamefont{J\"arvi
  et~al.}(2020{\natexlab{a}})\citenamefont{J\"arvi, Rinke, and
  Todorovi\'c}}]{Jaervi2020}
\bibinfo{author}{\bibfnamefont{J.}~\bibnamefont{J\"arvi}},
  \bibinfo{author}{\bibfnamefont{P.}~\bibnamefont{Rinke}}, \bibnamefont{and}
  \bibinfo{author}{\bibfnamefont{M.}~\bibnamefont{Todorovi\'c}},
  \bibinfo{journal}{Beilstein J. Nanotechnol.} \textbf{\bibinfo{volume}{11}},
  \bibinfo{pages}{1577} (\bibinfo{year}{2020}{\natexlab{a}}).

\bibitem[{\citenamefont{J\"arvi
  et~al.}(2020{\natexlab{b}})\citenamefont{J\"arvi, Alldritt, Krej\v{c}\'i,
  Todorovi\'c, Liljeroth, and Rinke}}]{Jaervi2021}
\bibinfo{author}{\bibfnamefont{J.}~\bibnamefont{J\"arvi}},
  \bibinfo{author}{\bibfnamefont{B.}~\bibnamefont{Alldritt}},
  \bibinfo{author}{\bibfnamefont{O.}~\bibnamefont{Krej\v{c}\'i}},
  \bibinfo{author}{\bibfnamefont{M.}~\bibnamefont{Todorovi\'c}},
  \bibinfo{author}{\bibfnamefont{P.}~\bibnamefont{Liljeroth}},
  \bibnamefont{and} \bibinfo{author}{\bibfnamefont{P.}~\bibnamefont{Rinke}},
  \bibinfo{journal}{Adv. Funct. Mater.} \textbf{\bibinfo{volume}{31}},
  \bibinfo{pages}{2010853} (\bibinfo{year}{2020}{\natexlab{b}}).

\bibitem[{\citenamefont{Fangnon et~al.}(2022)\citenamefont{Fangnon, Dvorak,
  Havu, Todorovi\'c, Li, and Rinke}}]{Fangnon2022}
\bibinfo{author}{\bibfnamefont{A.}~\bibnamefont{Fangnon}},
  \bibinfo{author}{\bibfnamefont{M.}~\bibnamefont{Dvorak}},
  \bibinfo{author}{\bibfnamefont{V.}~\bibnamefont{Havu}},
  \bibinfo{author}{\bibfnamefont{M.}~\bibnamefont{Todorovi\'c}},
  \bibinfo{author}{\bibfnamefont{J.}~\bibnamefont{Li}}, \bibnamefont{and}
  \bibinfo{author}{\bibfnamefont{P.}~\bibnamefont{Rinke}},
  \bibinfo{journal}{ACS Appl. Mater. Interfaces} \textbf{\bibinfo{volume}{14}},
  \bibinfo{pages}{12758} (\bibinfo{year}{2022}).

\bibitem[{\citenamefont{Chen et~al.}(2022)\citenamefont{Chen, Dong, Li, Zhu,
  Xu, Pan, Xu, Xi, Jiao, Hou et~al.}}]{ChenJ2022b}
\bibinfo{author}{\bibfnamefont{J.}~\bibnamefont{Chen}},
  \bibinfo{author}{\bibfnamefont{H.}~\bibnamefont{Dong}},
  \bibinfo{author}{\bibfnamefont{J.}~\bibnamefont{Li}},
  \bibinfo{author}{\bibfnamefont{X.}~\bibnamefont{Zhu}},
  \bibinfo{author}{\bibfnamefont{J.}~\bibnamefont{Xu}},
  \bibinfo{author}{\bibfnamefont{F.}~\bibnamefont{Pan}},
  \bibinfo{author}{\bibfnamefont{R.}~\bibnamefont{Xu}},
  \bibinfo{author}{\bibfnamefont{J.}~\bibnamefont{Xi}},
  \bibinfo{author}{\bibfnamefont{B.}~\bibnamefont{Jiao}},
  \bibinfo{author}{\bibfnamefont{X.}~\bibnamefont{Hou}}, \bibnamefont{et~al.},
  \bibinfo{journal}{ACS Energy Lett.} \textbf{\bibinfo{volume}{7}},
  \bibinfo{pages}{3685} (\bibinfo{year}{2022}).

\bibitem[{\citenamefont{Zhang et~al.}()\citenamefont{Zhang, Liao, Chen, Li, Xu,
  Wei, Du, Wang, Liu, Deng et~al.}}]{ZhangC2022}
\bibinfo{author}{\bibfnamefont{C.}~\bibnamefont{Zhang}},
  \bibinfo{author}{\bibfnamefont{Q.}~\bibnamefont{Liao}},
  \bibinfo{author}{\bibfnamefont{J.}~\bibnamefont{Chen}},
  \bibinfo{author}{\bibfnamefont{B.}~\bibnamefont{Li}},
  \bibinfo{author}{\bibfnamefont{C.}~\bibnamefont{Xu}},
  \bibinfo{author}{\bibfnamefont{K.}~\bibnamefont{Wei}},
  \bibinfo{author}{\bibfnamefont{G.}~\bibnamefont{Du}},
  \bibinfo{author}{\bibfnamefont{Y.}~\bibnamefont{Wang}},
  \bibinfo{author}{\bibfnamefont{D.}~\bibnamefont{Liu}},
  \bibinfo{author}{\bibfnamefont{J.}~\bibnamefont{Deng}}, \bibnamefont{et~al.},
  \bibinfo{journal}{Adv. Mater.} pp. \bibinfo{pages}{(in press,
  doi:10.1002/adma.202209422)} (????).

\bibitem[{\citenamefont{Stuke et~al.}(2021)\citenamefont{Stuke, Rinke, and
  Todorovi\'c}}]{Stuke2021}
\bibinfo{author}{\bibfnamefont{A.}~\bibnamefont{Stuke}},
  \bibinfo{author}{\bibfnamefont{P.}~\bibnamefont{Rinke}}, \bibnamefont{and}
  \bibinfo{author}{\bibfnamefont{M.}~\bibnamefont{Todorovi\'c}},
  \bibinfo{journal}{Mach. Learn.: Sci. Technol.} \textbf{\bibinfo{volume}{2}},
  \bibinfo{pages}{035022} (\bibinfo{year}{2021}).

\bibitem[{\citenamefont{Laakso et~al.}(2022)\citenamefont{Laakso, Todorovi\'c,
  Li, Zhang, and Rinke}}]{Laakso2022}
\bibinfo{author}{\bibfnamefont{J.}~\bibnamefont{Laakso}},
  \bibinfo{author}{\bibfnamefont{M.}~\bibnamefont{Todorovi\'c}},
  \bibinfo{author}{\bibfnamefont{J.}~\bibnamefont{Li}},
  \bibinfo{author}{\bibfnamefont{G.-X.} \bibnamefont{Zhang}}, \bibnamefont{and}
  \bibinfo{author}{\bibfnamefont{P.}~\bibnamefont{Rinke}},
  \bibinfo{journal}{Phys. Rev. Materials} \textbf{\bibinfo{volume}{6}},
  \bibinfo{pages}{113801} (\bibinfo{year}{2022}).

\bibitem[{\citenamefont{Sun et~al.}(2021)\citenamefont{Sun, Tiihonen, Oviedo,
  Liu, Thapa, Zhao, Hartono, Goyal, Heumueller, Batali et~al.}}]{SunS2021}
\bibinfo{author}{\bibfnamefont{S.}~\bibnamefont{Sun}},
  \bibinfo{author}{\bibfnamefont{A.}~\bibnamefont{Tiihonen}},
  \bibinfo{author}{\bibfnamefont{F.}~\bibnamefont{Oviedo}},
  \bibinfo{author}{\bibfnamefont{Z.}~\bibnamefont{Liu}},
  \bibinfo{author}{\bibfnamefont{J.}~\bibnamefont{Thapa}},
  \bibinfo{author}{\bibfnamefont{Y.}~\bibnamefont{Zhao}},
  \bibinfo{author}{\bibfnamefont{N.~T.~P.} \bibnamefont{Hartono}},
  \bibinfo{author}{\bibfnamefont{A.}~\bibnamefont{Goyal}},
  \bibinfo{author}{\bibfnamefont{T.}~\bibnamefont{Heumueller}},
  \bibinfo{author}{\bibfnamefont{C.}~\bibnamefont{Batali}},
  \bibnamefont{et~al.}, \bibinfo{journal}{Matter} \textbf{\bibinfo{volume}{4}},
  \bibinfo{pages}{1305} (\bibinfo{year}{2021}).

\bibitem[{\citenamefont{L\"ofgren et~al.}(2022)\citenamefont{L\"ofgren,
  Tarasov, Koitto, Rinke, Balakshin, and Todorovi\'c}}]{Loefgren2022}
\bibinfo{author}{\bibfnamefont{J.}~\bibnamefont{L\"ofgren}},
  \bibinfo{author}{\bibfnamefont{D.}~\bibnamefont{Tarasov}},
  \bibinfo{author}{\bibfnamefont{T.}~\bibnamefont{Koitto}},
  \bibinfo{author}{\bibfnamefont{P.}~\bibnamefont{Rinke}},
  \bibinfo{author}{\bibfnamefont{M.}~\bibnamefont{Balakshin}},
  \bibnamefont{and}
  \bibinfo{author}{\bibfnamefont{M.}~\bibnamefont{Todorovi\'c}},
  \bibinfo{journal}{ACS Sustain. Chem. Eng.} \textbf{\bibinfo{volume}{10}},
  \bibinfo{pages}{9469} (\bibinfo{year}{2022}).

\bibitem[{\citenamefont{Sim et~al.}(2019)\citenamefont{Sim, Jun, Bang, Kamioka,
  Kim, Hiramatsu, and Hosono}}]{SimK2019}
\bibinfo{author}{\bibfnamefont{K.}~\bibnamefont{Sim}},
  \bibinfo{author}{\bibfnamefont{T.}~\bibnamefont{Jun}},
  \bibinfo{author}{\bibfnamefont{J.}~\bibnamefont{Bang}},
  \bibinfo{author}{\bibfnamefont{H.}~\bibnamefont{Kamioka}},
  \bibinfo{author}{\bibfnamefont{J.}~\bibnamefont{Kim}},
  \bibinfo{author}{\bibfnamefont{H.}~\bibnamefont{Hiramatsu}},
  \bibnamefont{and} \bibinfo{author}{\bibfnamefont{H.}~\bibnamefont{Hosono}},
  \bibinfo{journal}{Appl. Phys. Rev.} \textbf{\bibinfo{volume}{6}},
  \bibinfo{pages}{031402} (\bibinfo{year}{2019}).

\bibitem[{\citenamefont{Glazer}(1972)}]{Glazer1972}
\bibinfo{author}{\bibfnamefont{A.~M.} \bibnamefont{Glazer}},
  \bibinfo{journal}{Acta Crystallogr.} \textbf{\bibinfo{volume}{B28}},
  \bibinfo{pages}{3384} (\bibinfo{year}{1972}).

\bibitem[{\citenamefont{Chung et~al.}(2012)\citenamefont{Chung, Song, Im,
  Androulakis, Malliakas, Li, Freeman, Kenney, and Kanatzidis}}]{ChungI2012}
\bibinfo{author}{\bibfnamefont{I.}~\bibnamefont{Chung}},
  \bibinfo{author}{\bibfnamefont{J.-H.} \bibnamefont{Song}},
  \bibinfo{author}{\bibfnamefont{J.}~\bibnamefont{Im}},
  \bibinfo{author}{\bibfnamefont{J.}~\bibnamefont{Androulakis}},
  \bibinfo{author}{\bibfnamefont{C.~D.} \bibnamefont{Malliakas}},
  \bibinfo{author}{\bibfnamefont{H.}~\bibnamefont{Li}},
  \bibinfo{author}{\bibfnamefont{A.~J.} \bibnamefont{Freeman}},
  \bibinfo{author}{\bibfnamefont{J.~T.} \bibnamefont{Kenney}},
  \bibnamefont{and} \bibinfo{author}{\bibfnamefont{M.~G.}
  \bibnamefont{Kanatzidis}}, \bibinfo{journal}{J. Am. Chem. Soc.}
  \textbf{\bibinfo{volume}{134}}, \bibinfo{pages}{8579} (\bibinfo{year}{2012}).

\bibitem[{\citenamefont{Stoumpos et~al.}(2013)\citenamefont{Stoumpos,
  Malliakas, Peters, Liu, Sebastian, Im, Chasapis, Wibowo, Chung, Freeman
  et~al.}}]{Stoumpos2013b}
\bibinfo{author}{\bibfnamefont{C.~C.} \bibnamefont{Stoumpos}},
  \bibinfo{author}{\bibfnamefont{C.~D.} \bibnamefont{Malliakas}},
  \bibinfo{author}{\bibfnamefont{J.~A.} \bibnamefont{Peters}},
  \bibinfo{author}{\bibfnamefont{Z.}~\bibnamefont{Liu}},
  \bibinfo{author}{\bibfnamefont{M.}~\bibnamefont{Sebastian}},
  \bibinfo{author}{\bibfnamefont{J.}~\bibnamefont{Im}},
  \bibinfo{author}{\bibfnamefont{T.~C.} \bibnamefont{Chasapis}},
  \bibinfo{author}{\bibfnamefont{A.~C.} \bibnamefont{Wibowo}},
  \bibinfo{author}{\bibfnamefont{D.~Y.} \bibnamefont{Chung}},
  \bibinfo{author}{\bibfnamefont{A.~J.} \bibnamefont{Freeman}},
  \bibnamefont{et~al.}, \bibinfo{journal}{Cryst. Growth Des.}
  \textbf{\bibinfo{volume}{13}}, \bibinfo{pages}{2722} (\bibinfo{year}{2013}).

\bibitem[{\citenamefont{Gehrmann and Egger}(2019)}]{Gehrmann2019}
\bibinfo{author}{\bibfnamefont{C.}~\bibnamefont{Gehrmann}} \bibnamefont{and}
  \bibinfo{author}{\bibfnamefont{D.~A.} \bibnamefont{Egger}},
  \bibinfo{journal}{Nature Commun.} \textbf{\bibinfo{volume}{10}},
  \bibinfo{pages}{3141} (\bibinfo{year}{2019}).

\bibitem[{\citenamefont{Glazer}(1975)}]{Glazer1975}
\bibinfo{author}{\bibfnamefont{A.~M.} \bibnamefont{Glazer}},
  \bibinfo{journal}{Acta Crystallogr.} \textbf{\bibinfo{volume}{A31}},
  \bibinfo{pages}{756} (\bibinfo{year}{1975}).

\bibitem[{\citenamefont{Woodward}(1997{\natexlab{a}})}]{Woodward1997a}
\bibinfo{author}{\bibfnamefont{P.~M.} \bibnamefont{Woodward}},
  \bibinfo{journal}{Acta Crystallogr.} \textbf{\bibinfo{volume}{B53}},
  \bibinfo{pages}{32} (\bibinfo{year}{1997}{\natexlab{a}}).

\bibitem[{\citenamefont{Woodward}(1997{\natexlab{b}})}]{Woodward1997b}
\bibinfo{author}{\bibfnamefont{P.~M.} \bibnamefont{Woodward}},
  \bibinfo{journal}{Acta Crystallogr.} \textbf{\bibinfo{volume}{B53}},
  \bibinfo{pages}{44} (\bibinfo{year}{1997}{\natexlab{b}}).

\bibitem[{\citenamefont{Islam et~al.}(2013)\citenamefont{Islam, Rondinelli, and
  Spanier}}]{Islam13}
\bibinfo{author}{\bibfnamefont{M.~A.} \bibnamefont{Islam}},
  \bibinfo{author}{\bibfnamefont{J.~M.} \bibnamefont{Rondinelli}},
  \bibnamefont{and} \bibinfo{author}{\bibfnamefont{J.~E.}
  \bibnamefont{Spanier}}, \bibinfo{journal}{J. Phys.: Condens. Matter}
  \textbf{\bibinfo{volume}{25}}, \bibinfo{pages}{175902}
  (\bibinfo{year}{2013}).

\bibitem[{\citenamefont{Xie et~al.}(2020)\citenamefont{Xie, Zhang, Raza, Zhang,
  Chen, and Wang}}]{XieN2020}
\bibinfo{author}{\bibfnamefont{N.}~\bibnamefont{Xie}},
  \bibinfo{author}{\bibfnamefont{J.}~\bibnamefont{Zhang}},
  \bibinfo{author}{\bibfnamefont{S.}~\bibnamefont{Raza}},
  \bibinfo{author}{\bibfnamefont{N.}~\bibnamefont{Zhang}},
  \bibinfo{author}{\bibfnamefont{X.}~\bibnamefont{Chen}}, \bibnamefont{and}
  \bibinfo{author}{\bibfnamefont{D.}~\bibnamefont{Wang}}, \bibinfo{journal}{J.
  Phys.: Condens. Matter} \textbf{\bibinfo{volume}{32}},
  \bibinfo{pages}{315901} (\bibinfo{year}{2020}).

\bibitem[{\citenamefont{Rasmussen and Williams}(2006)}]{Rasmussen2006}
\bibinfo{author}{\bibfnamefont{C.~E.} \bibnamefont{Rasmussen}}
  \bibnamefont{and} \bibinfo{author}{\bibfnamefont{C.~K.~I.}
  \bibnamefont{Williams}}, \emph{\bibinfo{title}{Gaussian Processes for Machine
  Learning}} (\bibinfo{publisher}{The MIT Press, Cambridge, MA},
  \bibinfo{year}{2006}).

\bibitem[{\citenamefont{Gutman and Corander}(2016)}]{Gutman2016}
\bibinfo{author}{\bibfnamefont{M.~U.} \bibnamefont{Gutman}} \bibnamefont{and}
  \bibinfo{author}{\bibfnamefont{J.}~\bibnamefont{Corander}},
  \bibinfo{journal}{J. Mach. Learn. Res} \textbf{\bibinfo{volume}{17}},
  \bibinfo{pages}{1} (\bibinfo{year}{2016}).

\bibitem[{\citenamefont{Brochu et~al.}()\citenamefont{Brochu, Cora, and
  {de~Freitas}}}]{Brochu2010}
\bibinfo{author}{\bibfnamefont{E.}~\bibnamefont{Brochu}},
  \bibinfo{author}{\bibfnamefont{V.~M.} \bibnamefont{Cora}}, \bibnamefont{and}
  \bibinfo{author}{\bibfnamefont{N.}~\bibnamefont{{de~Freitas}}},
  \bibinfo{note}{{arXiv}:1012.2599 (2010)}.

\bibitem[{\citenamefont{Perdew et~al.}(2008)\citenamefont{Perdew, Ruzsinszky,
  Csonka, Vydrov, Scuseria, Constantin, Zhou, and Burke}}]{Perdew2008}
\bibinfo{author}{\bibfnamefont{J.~P.} \bibnamefont{Perdew}},
  \bibinfo{author}{\bibfnamefont{A.}~\bibnamefont{Ruzsinszky}},
  \bibinfo{author}{\bibfnamefont{G.~I.} \bibnamefont{Csonka}},
  \bibinfo{author}{\bibfnamefont{O.~A.} \bibnamefont{Vydrov}},
  \bibinfo{author}{\bibfnamefont{G.~E.} \bibnamefont{Scuseria}},
  \bibinfo{author}{\bibfnamefont{L.~A.} \bibnamefont{Constantin}},
  \bibinfo{author}{\bibfnamefont{X.}~\bibnamefont{Zhou}}, \bibnamefont{and}
  \bibinfo{author}{\bibfnamefont{K.}~\bibnamefont{Burke}},
  \bibinfo{journal}{Phys. Rev. Lett.} \textbf{\bibinfo{volume}{100}},
  \bibinfo{pages}{136406} (\bibinfo{year}{2008}).

\bibitem[{\citenamefont{Knuth et~al.}(2015)\citenamefont{Knuth, Carbogno,
  Atalla, Blum, and Scheffler}}]{Knuth15}
\bibinfo{author}{\bibfnamefont{F.}~\bibnamefont{Knuth}},
  \bibinfo{author}{\bibfnamefont{C.}~\bibnamefont{Carbogno}},
  \bibinfo{author}{\bibfnamefont{V.}~\bibnamefont{Atalla}},
  \bibinfo{author}{\bibfnamefont{V.}~\bibnamefont{Blum}}, \bibnamefont{and}
  \bibinfo{author}{\bibfnamefont{M.}~\bibnamefont{Scheffler}},
  \bibinfo{journal}{Comput. Phys. Commun.} \textbf{\bibinfo{volume}{190}},
  \bibinfo{pages}{33} (\bibinfo{year}{2015}).

\bibitem[{\citenamefont{Blum et~al.}(2009)\citenamefont{Blum, Gehrke, Hanke,
  Havu, Havu, Ren, Reuter, and Scheffler}}]{Blum09}
\bibinfo{author}{\bibfnamefont{V.}~\bibnamefont{Blum}},
  \bibinfo{author}{\bibfnamefont{R.}~\bibnamefont{Gehrke}},
  \bibinfo{author}{\bibfnamefont{F.}~\bibnamefont{Hanke}},
  \bibinfo{author}{\bibfnamefont{P.}~\bibnamefont{Havu}},
  \bibinfo{author}{\bibfnamefont{V.}~\bibnamefont{Havu}},
  \bibinfo{author}{\bibfnamefont{X.}~\bibnamefont{Ren}},
  \bibinfo{author}{\bibfnamefont{K.}~\bibnamefont{Reuter}}, \bibnamefont{and}
  \bibinfo{author}{\bibfnamefont{M.}~\bibnamefont{Scheffler}},
  \bibinfo{journal}{Comput. Phys. Commun.} \textbf{\bibinfo{volume}{180}},
  \bibinfo{pages}{2175} (\bibinfo{year}{2009}).

\bibitem[{\citenamefont{Havu et~al.}(2009)\citenamefont{Havu, Blum, Havu, and
  Scheffler}}]{HavuV09}
\bibinfo{author}{\bibfnamefont{V.}~\bibnamefont{Havu}},
  \bibinfo{author}{\bibfnamefont{V.}~\bibnamefont{Blum}},
  \bibinfo{author}{\bibfnamefont{P.}~\bibnamefont{Havu}}, \bibnamefont{and}
  \bibinfo{author}{\bibfnamefont{M.}~\bibnamefont{Scheffler}},
  \bibinfo{journal}{J. Comput. Phys.} \textbf{\bibinfo{volume}{228}},
  \bibinfo{pages}{8367} (\bibinfo{year}{2009}).

\bibitem[{\citenamefont{Ren et~al.}(2012)\citenamefont{Ren, Rinke, Blum,
  Wieferink, Tkatchenko, Sanfilippo, Reuter, and Scheffler}}]{RenX12}
\bibinfo{author}{\bibfnamefont{X.}~\bibnamefont{Ren}},
  \bibinfo{author}{\bibfnamefont{P.}~\bibnamefont{Rinke}},
  \bibinfo{author}{\bibfnamefont{V.}~\bibnamefont{Blum}},
  \bibinfo{author}{\bibfnamefont{J.}~\bibnamefont{Wieferink}},
  \bibinfo{author}{\bibfnamefont{A.}~\bibnamefont{Tkatchenko}},
  \bibinfo{author}{\bibfnamefont{A.}~\bibnamefont{Sanfilippo}},
  \bibinfo{author}{\bibfnamefont{K.}~\bibnamefont{Reuter}}, \bibnamefont{and}
  \bibinfo{author}{\bibfnamefont{M.}~\bibnamefont{Scheffler}},
  \bibinfo{journal}{New J. Phys.} \textbf{\bibinfo{volume}{14}},
  \bibinfo{pages}{053020} (\bibinfo{year}{2012}).

\bibitem[{\citenamefont{Levchenko et~al.}(2015)\citenamefont{Levchenko, Ren,
  Wieferink, Johanni, Rinke, Blum, and Scheffler}}]{Levchenko15}
\bibinfo{author}{\bibfnamefont{S.~V.} \bibnamefont{Levchenko}},
  \bibinfo{author}{\bibfnamefont{X.}~\bibnamefont{Ren}},
  \bibinfo{author}{\bibfnamefont{J.}~\bibnamefont{Wieferink}},
  \bibinfo{author}{\bibfnamefont{R.}~\bibnamefont{Johanni}},
  \bibinfo{author}{\bibfnamefont{P.}~\bibnamefont{Rinke}},
  \bibinfo{author}{\bibfnamefont{V.}~\bibnamefont{Blum}}, \bibnamefont{and}
  \bibinfo{author}{\bibfnamefont{M.}~\bibnamefont{Scheffler}},
  \bibinfo{journal}{Comput. Phys. Commun.} \textbf{\bibinfo{volume}{192}},
  \bibinfo{pages}{60} (\bibinfo{year}{2015}).

\bibitem[{\citenamefont{Yang et~al.}(2017)\citenamefont{Yang, Skelton,
  {da~Silva}, Frost, and Walsh}}]{YangRX2017}
\bibinfo{author}{\bibfnamefont{R.~X.} \bibnamefont{Yang}},
  \bibinfo{author}{\bibfnamefont{J.~M.} \bibnamefont{Skelton}},
  \bibinfo{author}{\bibfnamefont{E.~L.} \bibnamefont{{da~Silva}}},
  \bibinfo{author}{\bibfnamefont{J.~M.} \bibnamefont{Frost}}, \bibnamefont{and}
  \bibinfo{author}{\bibfnamefont{A.}~\bibnamefont{Walsh}}, \bibinfo{journal}{J.
  Phys. Chem. Lett.} \textbf{\bibinfo{volume}{8}}, \bibinfo{pages}{4720}
  (\bibinfo{year}{2017}).

\bibitem[{\citenamefont{Bokdam et~al.}(2017)\citenamefont{Bokdam, Lahnsteiner,
  Ramberger, Sch\"afer, and Kresse}}]{Bokdam2017}
\bibinfo{author}{\bibfnamefont{M.}~\bibnamefont{Bokdam}},
  \bibinfo{author}{\bibfnamefont{J.}~\bibnamefont{Lahnsteiner}},
  \bibinfo{author}{\bibfnamefont{B.}~\bibnamefont{Ramberger}},
  \bibinfo{author}{\bibfnamefont{T.}~\bibnamefont{Sch\"afer}},
  \bibnamefont{and} \bibinfo{author}{\bibfnamefont{G.}~\bibnamefont{Kresse}},
  \bibinfo{journal}{Phys. Rev. Lett.} \textbf{\bibinfo{volume}{119}},
  \bibinfo{pages}{145501} (\bibinfo{year}{2017}).

\bibitem[{\citenamefont{Seidu et~al.}(2021)\citenamefont{Seidu, Dvorak, Rinke,
  and Li}}]{Seidu2021a}
\bibinfo{author}{\bibfnamefont{A.}~\bibnamefont{Seidu}},
  \bibinfo{author}{\bibfnamefont{M.}~\bibnamefont{Dvorak}},
  \bibinfo{author}{\bibfnamefont{P.}~\bibnamefont{Rinke}}, \bibnamefont{and}
  \bibinfo{author}{\bibfnamefont{J.}~\bibnamefont{Li}}, \bibinfo{journal}{J.
  Chem. Phys.} \textbf{\bibinfo{volume}{154}}, \bibinfo{pages}{074712}
  (\bibinfo{year}{2021}).

\bibitem[{\citenamefont{van Lenthe et~al.}(1993)\citenamefont{van Lenthe,
  Baerends, and Sneijders}}]{vanLenthe93}
\bibinfo{author}{\bibfnamefont{E.}~\bibnamefont{van Lenthe}},
  \bibinfo{author}{\bibfnamefont{E.~J.} \bibnamefont{Baerends}},
  \bibnamefont{and} \bibinfo{author}{\bibfnamefont{J.~G.}
  \bibnamefont{Sneijders}}, \bibinfo{journal}{J. Chem. Phys.}
  \textbf{\bibinfo{volume}{99}}, \bibinfo{pages}{4597} (\bibinfo{year}{1993}).

\bibitem[{NoM()}]{NoMaD-BOSS_CPX}
\bibinfo{note}{See
  {\href{https://dx.doi.org/10.17172/NOMAD/2021.09.15-1}{https://dx.doi.org/10.17172/NOMAD/2021.09.15-1}}}.

\bibitem[{\citenamefont{Wang et~al.}(2019)\citenamefont{Wang, {Ibrahim~Dar},
  Ono, Zhang, Kan, Li, Zhang, Wang, Yang, Gao et~al.}}]{WangY2019}
\bibinfo{author}{\bibfnamefont{Y.}~\bibnamefont{Wang}},
  \bibinfo{author}{\bibfnamefont{M.}~\bibnamefont{{Ibrahim~Dar}}},
  \bibinfo{author}{\bibfnamefont{L.~K.} \bibnamefont{Ono}},
  \bibinfo{author}{\bibfnamefont{T.}~\bibnamefont{Zhang}},
  \bibinfo{author}{\bibfnamefont{M.}~\bibnamefont{Kan}},
  \bibinfo{author}{\bibfnamefont{Y.}~\bibnamefont{Li}},
  \bibinfo{author}{\bibfnamefont{L.}~\bibnamefont{Zhang}},
  \bibinfo{author}{\bibfnamefont{X.}~\bibnamefont{Wang}},
  \bibinfo{author}{\bibfnamefont{Y.}~\bibnamefont{Yang}},
  \bibinfo{author}{\bibfnamefont{X.}~\bibnamefont{Gao}}, \bibnamefont{et~al.},
  \bibinfo{journal}{Science} \textbf{\bibinfo{volume}{365}},
  \bibinfo{pages}{591} (\bibinfo{year}{2019}).

\bibitem[{\citenamefont{Hirotsu et~al.}(2013)\citenamefont{Hirotsu, Harada,
  Iizumi, and Gesi}}]{Hirotsu1974}
\bibinfo{author}{\bibfnamefont{S.}~\bibnamefont{Hirotsu}},
  \bibinfo{author}{\bibfnamefont{J.}~\bibnamefont{Harada}},
  \bibinfo{author}{\bibfnamefont{M.}~\bibnamefont{Iizumi}}, \bibnamefont{and}
  \bibinfo{author}{\bibfnamefont{K.}~\bibnamefont{Gesi}}, \bibinfo{journal}{J.
  Phys. Soc. Jpn.} \textbf{\bibinfo{volume}{37}}, \bibinfo{pages}{1393}
  (\bibinfo{year}{2013}).

\bibitem[{\citenamefont{Eperon et~al.}(2014)\citenamefont{Eperon, Stranks,
  Menelaou, Johnston, Herz, and Snaith}}]{Eperon14}
\bibinfo{author}{\bibfnamefont{G.~E.} \bibnamefont{Eperon}},
  \bibinfo{author}{\bibfnamefont{S.~D.} \bibnamefont{Stranks}},
  \bibinfo{author}{\bibfnamefont{C.}~\bibnamefont{Menelaou}},
  \bibinfo{author}{\bibfnamefont{M.~B.} \bibnamefont{Johnston}},
  \bibinfo{author}{\bibfnamefont{L.~M.} \bibnamefont{Herz}}, \bibnamefont{and}
  \bibinfo{author}{\bibfnamefont{H.~J.} \bibnamefont{Snaith}},
  \bibinfo{journal}{Energy Environ. Sci.} \textbf{\bibinfo{volume}{7}},
  \bibinfo{pages}{982} (\bibinfo{year}{2014}).

\bibitem[{\citenamefont{Eperon et~al.}(2015)\citenamefont{Eperon, Patern\`o,
  Sutton, Zampetti, Haghighirad, Cacialli, and Snaith}}]{Eperon15}
\bibinfo{author}{\bibfnamefont{G.~E.} \bibnamefont{Eperon}},
  \bibinfo{author}{\bibfnamefont{G.~M.} \bibnamefont{Patern\`o}},
  \bibinfo{author}{\bibfnamefont{R.~J.} \bibnamefont{Sutton}},
  \bibinfo{author}{\bibfnamefont{A.}~\bibnamefont{Zampetti}},
  \bibinfo{author}{\bibfnamefont{A.~A.} \bibnamefont{Haghighirad}},
  \bibinfo{author}{\bibfnamefont{F.}~\bibnamefont{Cacialli}}, \bibnamefont{and}
  \bibinfo{author}{\bibfnamefont{H.}~\bibnamefont{Snaith}},
  \bibinfo{journal}{J. Mater. Chem. A} \textbf{\bibinfo{volume}{3}},
  \bibinfo{pages}{19688} (\bibinfo{year}{2015}).

\bibitem[{\citenamefont{Mannino et~al.}(2020)\citenamefont{Mannino, Deretzis,
  Smecca, {La~Magna}, Alberti, Ceratti, and Cahen}}]{Mannino2020}
\bibinfo{author}{\bibfnamefont{G.}~\bibnamefont{Mannino}},
  \bibinfo{author}{\bibfnamefont{I.}~\bibnamefont{Deretzis}},
  \bibinfo{author}{\bibfnamefont{E.}~\bibnamefont{Smecca}},
  \bibinfo{author}{\bibfnamefont{A.}~\bibnamefont{{La~Magna}}},
  \bibinfo{author}{\bibfnamefont{A.}~\bibnamefont{Alberti}},
  \bibinfo{author}{\bibfnamefont{D.}~\bibnamefont{Ceratti}}, \bibnamefont{and}
  \bibinfo{author}{\bibfnamefont{D.}~\bibnamefont{Cahen}}, \bibinfo{journal}{J.
  Phys. Chem. Lett.} \textbf{\bibinfo{volume}{11}}, \bibinfo{pages}{2490}
  (\bibinfo{year}{2020}).

\bibitem[{\citenamefont{Shockley and Queisser}(1961)}]{Shockley61}
\bibinfo{author}{\bibfnamefont{W.}~\bibnamefont{Shockley}} \bibnamefont{and}
  \bibinfo{author}{\bibfnamefont{H.-J.} \bibnamefont{Queisser}},
  \bibinfo{journal}{J. Appl. Phys.} \textbf{\bibinfo{volume}{32}},
  \bibinfo{pages}{510} (\bibinfo{year}{1961}).

\end{thebibliography}

\onecolumngrid

\clearpage

\pagestyle{empty}

{\centering
\large \textbf{Supporting Information}

}

\clearpage
\includegraphics[trim=0.9in 0in 0.9in 1in,clip=true,width=\textwidth,page=2]{./supporting_information.pdf}
\clearpage
\includegraphics[trim=0.9in 0in 0.9in 1in,clip=true,width=\textwidth,page=3]{./supporting_information.pdf}
\clearpage
\includegraphics[trim=0.9in 0in 0.9in 1in,clip=true,width=\textwidth,page=4]{./supporting_information.pdf}
\clearpage
\includegraphics[trim=0.9in 0in 0.9in 1in,clip=true,width=\textwidth,page=5]{./supporting_information.pdf}
\clearpage
\includegraphics[trim=0.9in 0in 0.9in 1in,clip=true,width=\textwidth,page=6]{./supporting_information.pdf}
\clearpage
\includegraphics[trim=0.9in 0in 0.9in 1in,clip=true,width=\textwidth,page=7]{./supporting_information.pdf}
\clearpage
\includegraphics[trim=0.9in 0in 0.9in 1in,clip=true,width=\textwidth,page=8]{./supporting_information.pdf}
\clearpage
\includegraphics[trim=0.9in 0in 0.9in 1in,clip=true,width=\textwidth,page=9]{./supporting_information.pdf}
\clearpage
\includegraphics[trim=0.9in 0in 0.9in 1in,clip=true,width=\textwidth,page=10]{./supporting_information.pdf}
\clearpage
\includegraphics[trim=0.9in 0in 0.9in 1in,clip=true,width=\textwidth,page=11]{./supporting_information.pdf}
\clearpage
\includegraphics[trim=0.9in 0in 0.9in 1in,clip=true,width=\textwidth,page=12]{./supporting_information.pdf}
\clearpage
\includegraphics[trim=0.9in 0in 0.9in 1in,clip=true,width=\textwidth,page=13]{./supporting_information.pdf}

\end{document}